\documentclass[titlepage,12pt,twoside,table]{article}

\usepackage[dvips]{graphicx}
\usepackage[myheadings]{fullpage}

\usepackage{pgfplots}
\usepackage{ifthen}
\usepackage{indentfirst}
\usepackage{graphicx}
\usepackage{ragged2e}
\usepackage{bm}
\usepackage{float}
\usepackage{placeins}
\usepackage{amssymb}
\usepackage{amsmath}
\usepackage{soul}
\usepackage{placeins}
\usepackage{mathtools}
\usepackage{dcolumn}
\newcolumntype{.}{D{.}{.}{1}}
\newcolumntype{,}{D{,}{,}{1}}
\usepackage{tikz}
\usepackage{multirow}
\usepackage[round,comma]{natbib}
\usepackage[thinlines]{easytable}
\usepackage{url}
\usepackage{authblk}
\edef\restoreparindent{\parindent=\the\parindent\relax}
\usepackage[parfill]{parskip}
\restoreparindent

\pgfplotsset{compat=1.14}

\DeclareMathOperator{\Var}{Var}
\DeclareMathOperator{\cov}{cov}

\DeclareMathOperator{\PP}{\mathbb{P}}

\usepackage{adjustbox}
\usepackage{fancyvrb}
\title{A Dyadic IRT Model}
\author[1]{Brian Gin}
\author[4]{Nicholas Sim}
\author[2,3,4]{Anders Skrondal}
\author[4]{Sophia Rabe-Hesketh}
\affil[1]{University of California, San Francisco}
\affil[2]{Norwegian Institute of Public Health}
\affil[3]{University of Oslo}
\affil[4]{University of California, Berkeley}
\date{}

\begin{document}

\maketitle

\begin{abstract}
We propose a dyadic Item Response Theory (dIRT) model for measuring interactions of pairs of individuals when the responses to items represent the actions (or behaviors, perceptions, etc.) of each individual (actor) made within the context of a dyad formed with another individual (partner). Examples of its use include the assessment of collaborative problem solving, or the evaluation of intra-team dynamics. The dIRT model generalizes both Item Response Theory (IRT) models for measurement and the Social Relations Model (SRM) for dyadic data. The responses of an actor when paired with a partner are modeled as a function of not only the actor's inclination to act and the partner's tendency to elicit that action, but also the unique relationship of the pair, represented by two directional, possibly correlated, interaction latent variables. Generalizations are discussed, such as accommodating triads or larger groups. Estimation is performed using Markov-chain Monte Carlo implemented in \texttt{Stan}, making it straightforward to extend the dIRT model in various ways. Specifically, we show how the basic dIRT model can be extended to accommodate latent regressions, multilevel settings with cluster-level random effects, as well as joint modeling of dyadic data and a distal outcome. A simulation study demonstrates that estimation performs well. We apply our proposed approach to speed-dating data and find new evidence of pairwise interactions between participants, describing a mutual attraction that is inadequately characterized by individual properties alone.

\end{abstract}

\section{Introduction}
The study of how individuals interact within a group has been and continues to be of interest to researchers in the behavioral sciences.  Even in this setting, the majority of statistical models focus primarily on how each individual behaves isolated from the influences of other group members.  However, one model developed to handle the simplest case of two individuals interacting in a dyad is the Social Relations Model (SRM) \citep[e.g.,][]{Warner:79,Kenny84,Kenny06}.  Here, the ways one individual (often called an actor or perceiver) of a dyad behaves when paired with the other (often called the partner or target) and vice-versa are analyzed to infer individual-level and dyad-level effects.  The behavior of the actors can be directed towards the partner (e.g., an individual's perception of the partner's attractiveness) or undirected (e.g., the number of times an individual takes the lead in a collaborative problem solving task), and can be measured during or after socially interacting with the partner.  Compared with traditional ``isolated'' models, the innovative SRM considers both members of the dyad as contributors to the eventual observed behavior. The SRM model has been most often used in social psychology \citep[e.g.,][]{kennyk:94}, but is increasingly being used in other fields. A diverse set of examples include relationships in pharmacy and therapeutics hospital-committee decision-making \citep{bagozzi:05}, social media ties among basketball teammates \citep{koster:2018}, and militarized interstate disputes \citep{dorff:2013}.

In the original formulation of the SRM, the specific behavior of an actor when paired with a partner depends on a composite dyad-level latent trait that can be decomposed into three parts: (i) an individual-level latent trait reflecting a general inclination of the actor to behave in a certain way when paired with a partner, (ii) an individual-level latent trait reflecting the general tendency of the partner to elicit such a behavior, and (iii) a dyad-level latent trait that characterizes the effect of the unique (directed) relationship between both parties on the behavior of the actor that is independent of the two individual-level latent traits \citep{Back}.  More concretely, if one is interested in the level of physical attraction of an actor towards a partner, then the three components reflect (i) how, on average, an actor tends to find others attractive, (ii) how, on average, the partner tends to be found attractive, and how (iii) the actor uniquely finds the partner attractive.  As a result of this formulation, the SRM is identifiable only if individuals belong to multiple pairs.

While the SRM is a useful tool in the analysis of dyads, it has not yet been extended for the case where a set of behaviors or responses of an actor can be viewed as measuring a latent variable, such as the actor's perception of or disposition towards a partner. Multivariate SRM~\citep[e.g.,][]{kenny:94, card:2008, nestler:2018} accommodates multiple measures, but it effectively corresponds to a set of univariate SRMs with additional correlations of individual-level and dyad-level latent traits across measures. When there are more than two or three measures, the multivariate SRM has an abundance of cross-variable correlations that are not easy to interpret. More importantly, multivariate SRMs do not provide a means for predicting or scoring actor, partner and dyad effects on an underlying latent trait. With existing methodology, a better alternative would be to specify a univariate SRM for some summary of the measures, such as a sum-score or mean. However, this could result in a loss of information analogous to educational testing where the scores on different items of the test are sometimes summed up, and only the sum score is used. Our proposed dyadic Item Response Theory (dIRT) model therefore incorporates an Item Response Theory (IRT) model. Advantages include having the ability to account for differences in item difficulty, allowing for missing responses in subsets of items (under the Missing-at-Random assumption), and having individualized standard errors of the latent trait scores~\citep[e.g.,][]{Embretson}.

IRT is the standard approach for  modeling the relationship between the latent traits  of individuals and their responses to a set of items in educational testing.  There are a variety of IRT models that may differ, among other things, in terms of the numbers of parameters in the model, the type of link function used, or the approach taken (e.g., confirmatory or exploratory) \citep[e.g.,][]{Linden}.  However, existing models treat the latent trait as a property of the individuals who responded to the items, and perhaps an external party like a rater, but do not include a unique interaction between individuals in a dyad.  That is, traditional IRT can be used to model the behavior of an actor when paired with a partner as a function of the items/stimuli, the  actor's tendency to behave in a certain way and perhaps the partner's tendency to elicit the behavior, but does not accommodate the unique dyadic effect due to both individuals interacting in a social setting.  Thus, if individuals interact with one another, and the manner and effect of this interaction is of interest, then existing IRT models are not useful.

Although SRM and IRT models each have limitations that could be overcome by the other, there is, to our knowledge, no prior work on integrating the models. Only two related cases appear to exists: \citet{Alex} extended the Actor-Partner Interdependence Model (APIM)  and Common Fate Model (CFM) that we describe in Section~\ref{sec:other} to work within an IRT framework. While these models relax the condition that only an individual's latent ability affects the individual's response to an item, neither of them models the dyadic interaction as a latent trait of the dyad.  Furthermore, the APIM and CFM are limited to a dyadic design where each individual is paired with only one other partner whereas the SRM handles the case where individuals belong to multiple pairs \citep{Kenny06}.

Our contributions include the following.  First, we describe our proposed dIRT model that incorporates the key features of both the SRM and IRT. The model includes individual and dyad-level latent traits and corresponding variance and covariance parameters afforded by the SRM, while retaining all the important measurement properties of IRT. We also indicate how the model can be extended to larger groupings than dyads, such as triads. Second, we provide a literature review of related classes of models and discuss data designs and conditions for identifiability. Importantly, unlike the SRM, the dIRT model is identified for cross-sectional data. Third,  we extend the basic dIRT model to let the latent traits affect a distal outcome and depend on observed covariates and cluster-level random effects.  Finally, we demonstrate the practical utility of the model by applying it to a speed dating dataset and making \texttt{Stan} code available, together with a case-study explaining the code. While univariate SRMs for one Likert scale item at a time, treated as continuous, have been applied to speed-dating data~\citep[e.g.,][]{ackerman:2015}, our multivariate model  accommodates the ordinal nature of the responses and allows estimation of
the unique interaction variance separate from the error variance. We hope that our contributions will inspire researchers to collect and analyze dyadic data in new settings.

The structure of the paper is as follows.  In Section \ref{sec:dIRT}, we introduce the basic dIRT model, discuss data design and identification, propose various extensions of the basic model, and provide a review of related models.  We present a Markov-chain Monte Carlo approach to estimating the model in Section \ref{sec:est}, using \texttt{Stan} for estimation.  In Section \ref{sec:app}, we apply our model and estimation method to a publicly available speed-dating dataset.  In Section \ref{sec:sim} we conduct a simulation study to evaluate the performance of our estimator under a variety of conditions.  Finally, we make some concluding remarks in Section \ref{sec:conc}.

\section{Dyadic Item Response Theory (dIRT)\label{sec:dIRT}}
\subsection{Basic dIRT Model}

In a social setting where groups of individuals interact, it is likely that the behavior of individual $a \in \{1, 2, \ldots, n\}$ (called the actor) in group $g$ is affected not only by his/her own latent traits, but also those of the individuals he/she interacts with.  Additionally, there could also be a ``unique'' component attributable to the specific composition of the group that could affect the actor's behavior above and beyond the effects at the individual level.  We can extend any IRT model to deal with such a setting by replacing the latent trait $\theta_a$ of individual $a$, with a composite latent trait $\theta_{a, g}$ of individual $a$ in the context of group $g$ of size $n$:
\begin{equation}
\label{generallinear}
\theta_{a,g} \ \equiv \ \alpha_a + \sum_{\substack{j = 1\\j \ne a}}^n \beta_j + \sum_{k \in K} \gamma_{a, g(k)}.
\end{equation}
Here, $\alpha_a$ represents the inclination of the actor to behave in a certain way, $\beta_j$ represents the tendency of another member $j$ of the group to elicit the behavior, and $\gamma_{a,g(k)}$ represents the unique way members of subgroup $g(k)$ interacted to elicit the behavior from actor $a$.  The last sum above is over all possible subgroups $g(k)$ of sizes 1 to $n-1$ excluding the actor. (The index set is defined as $K := \{A\subseteq \{1, 2, \ldots, n\}\setminus\{a\} \mid |A| \ge 1\}$, i.e., the set of all subsets of $\{1, 2, \ldots, n\}\setminus\{a\}$ except the empty set).  Note that $\gamma_{a,g(k)}$ includes not only physical interactions between actor $a$ and the other members of the group, but also how the behavior of actor $a$ is altered by the mere presence of the rest of the group.  For example, in a collaborative problem solving task, $\alpha_a$ could represent the inclination of  actor $a$ to be vocal, $\beta_j$  how much  partner $j$ tends to elicit opinions from actors, and $\gamma_{a,g(k)}$ 
how vocal the actor is due to the composition of the group $g(k)$.  In practice, it may not be necessary to include anything more than pairwise and possibly three-way interactions.

To simplify notation, in the rest of the paper, we focus on the case when $n = 2$ as it is clear how the model can be extended when working with larger group sizes.  In this dyadic setting, for actor $a$ and partner $p$, the composite latent trait is modeled as
\begin{equation*}
\theta_{a,p} \ \equiv \ \alpha_a + \beta_p + \gamma_{a,p}.
\end{equation*}
Unlike (\ref{generallinear}) where the composite latent variable $\theta$, and in particular the dyad-level latent trait $\gamma$, are indexed by the actor and the group, we can instead index $\theta$ and $\gamma$ by both individuals $a$ and $p$ since the index set $K$ reduces to the singleton set $\{\{p\}\}$.  Here, $\alpha_a$ is the actor latent trait (sometimes called actor effect), $\beta_p$ the partner latent trait (sometimes called partner effect), and $\gamma_{a, p}$ the dyadic latent trait (sometimes called interaction or relationship effect) which represents the unique contribution of pairing actor $a$ with partner $p$ to the behavior of the actor.  Note that $\gamma_{a, p}$ is not assumed to be identical to $\gamma_{p, a}$ when the roles of actor and partner are reversed.

We could consider any traditional IRT model for measuring $\theta_{a,p}$. The model for response $y_{a,p,i}$ to item $i$ by actor $a$, when paired with the partner $p$, is of the form
\begin{equation*}
g(\PP(y_{a, p, i}=j \mid \theta_{a, p}, \bm{\xi}_{i,j})) \ = \ f(\theta_{a, p}, \bm{\xi}_{i,j})
\end{equation*}
for some link function $g(\cdot)$, item parameters $\bm{\xi}_{i,j}$, and functional form $f(\cdot)$.  
For instance, for ordinal responses we can obtain the standard partial credit model \citep{Masters} by using the adjacent-category logit link, letting $\bm{\xi}_{i,j}$ represent (unidimensional) step difficulty parameters, and taking $f(\cdot)$
to be the identity function. If item $i$ has $m_i$ categories (from 0 to $m_i - 1$), the model becomes
\begin{equation}
\label{dirt}
\log \left( \frac{\PP_{\textnormal{PCM}}(y_{a, p, i}=j \mid \theta_{a, p}, \delta_{i,j})}{\PP_{\textnormal{PCM}}(y_{a, p, i}=j-1 \mid \theta_{a, p}, \delta_{i,j})} \right) \ = \ \theta_{a, p} - \delta_{i, j} \ \equiv \ (\alpha_a + \beta_p + \gamma_{a,p}) - \delta_{i, j},
\end{equation}
subject to the constraint that $\sum_{j=0}^{m_i-1} \PP_{\textnormal{PCM}}(y_{a, p, i}=j \mid \theta_{a, p},  \delta_{i, j}) = 1$, where $j \in \{1, 2, \ldots, m_i-1\}$, and $\delta_{i, j}$ are item step difficulties.  Note that we condition on $\delta_{i,j}$ because we will adopt a (pragmatic) Bayesian perspective (see Section 3).

In the dIRT model, we assume that the latent traits (or random effects) have bivariate normal distributions:
\begin{align}
\left[ \begin{array}{c}
\alpha_a \\
\beta_a \\
\end{array} \right]
& \ \sim \
\textnormal{N}\left(
\left[\begin{array}{c}
\mu_\alpha \\
\mu_\beta \\
\end{array} \right]
,
\left[ \begin{array}{cc}
\sigma_\alpha^2 & \rho_{\alpha\beta}\sigma_\alpha\sigma_\beta \\
\rho_{\alpha\beta}\sigma_\alpha\sigma_\beta  & \sigma_\beta^2 \\
\end{array} \right]
\right), \nonumber \\ 
\left[ \begin{array}{c}
\gamma_{a, p} \\
\gamma_{p, a} \\
\end{array} \right]
& \ \sim \
\textnormal{N}\left(
\left[ \begin{array}{c}
\mu_\gamma \\
\mu_\gamma \\
\end{array} \right]
,
\left[ \begin{array}{cc}
\sigma_\gamma^2 & \rho_{\gamma}\sigma_\gamma^2 \\
\rho_{\gamma}\sigma_\gamma^2  & \sigma_\gamma^2 \\
\end{array} \right]
\right).
\label{hyperparameters}
\end{align}
The parameters are (i) the variances $\sigma_\alpha^2$, $\sigma_\beta^2$, and $\sigma_\gamma^2$ of the individual and dyad latent traits, (ii), the expectations $\mu_\alpha$, $\mu_\beta$, and $\mu_\gamma$ of each of the individual and dyadic latent traits, and (iii) the correlations $\rho_{\alpha\beta}$ and $\rho_\gamma$.

The individual-level correlation $\rho_{\alpha\beta}$ (sometimes called the general or individual reciprocity) relates the tendency of an individual to behave in a certain way (i.e., $\alpha_a$ or $\alpha_p$) to that same individual's tendency to elicit the behavior from his/her partner (i.e., $\beta_a$ or $\beta_p$). The dyad-level correlation $\rho_\gamma$ (sometimes called dyadic reciprocity) relates the two (directed) latent traits of each dyad (i.e., $\gamma_{a, p}$ and $\gamma_{p, a}$) to each other.

We will extend the dIRT model in Section~\ref{sec:extdIRT} after discussing data design and identification issues that will motivate and justify some of the extensions.

\subsection{Data Design and Identification\label{sec:design}}

The dIRT model has five variance-covariance parameters for the individual and dyadic latent traits that imply five ``reduced-form parameters'' for the composite latent trait: one constant variance, $\Var(\theta_{a,p})=\sigma_\alpha^2+\sigma_\beta^2+\sigma_\gamma^2$, and four distinct non-zero covariances, $\cov(\theta_{a,p},\theta_{p,a})=2\rho_{\alpha\beta}\sigma_{\alpha}\sigma_{\beta}+\rho_{\gamma}\sigma_\gamma^2$), $\cov(\theta_{a,p},\theta_{a,q})=\sigma_\alpha^2$, $\cov(\theta_{a,p},\theta_{b,p})=\sigma_\beta^2$, and $\cov(\theta_{a,p},\theta_{b,a})=\rho_{\alpha\beta}\sigma_\alpha\sigma_\beta$ (where $a$, $p$, $b$, $q$ are all different individuals). It is straightforward to find unique solutions for the five variance-covariance parameters from the five equations above, showing that the they are identified if the reduced-form parameters (variance and covariances) are identified. 

The reduced-form parameters are identified if all the pairs of dyads involved in the covariances exist, i.e., actor/partner role reversal  (sometimes referred to as ``reciprocals'') must occur to identify $\cov(\theta_{a,p},\theta_{p,a})$ and it must be possible to belong to more than one dyad. Specifically, it must be possible for actors to be paired with several partners to identify $\cov(\theta_{a,p},\theta_{a,q})$, for partners to be paired with several actors to identify $\cov(\theta_{a,p},\theta_{b,p})$,  and for an actor paired
with a partner $p$ to also occur in a dyad as a partner of an actor $b\neq p$ to identify $\cov(\theta_{a,p},\theta_{b,a})$. 
It is necessary to set some mean parameters and/or step difficulty parameters to constants for identification. Here, we set the expectations of the latent traits to zero $(\mu_\alpha = \mu_\beta = \mu_\gamma = 0)$, and allow the item step difficulties $\delta_{i, j}$ to be unconstrained (anchoring on latent trait scores instead of item difficulties), except that $\delta_{i,0} = 0$.

We have implicitly assumed that dyads and individuals within dyads are exchangeable by restricting the mean of $\theta_{a,p}$ to be constant (set to 0 to identify the step-difficulties) and allowing for only  five distinct second-order reduced form parameters (one variance and four covariances), i.e., by assuming that the variance is constant and that covariances between dyadic composite latent variables depend only on the actor/partner roles of the individuals that are present in both dyads. The corresponding five parameters $\sigma_\alpha^2$, $\sigma_\beta^2$, $\sigma_\gamma^2$, $\rho_{\alpha\beta}$, and $\rho_{\gamma}$ enforce no other constraints besides exchangeability and positive semi-definiteness. \citet{li:2002} make the point, for a regular SRM, that the model is in that sense justified by exchangeability.

When dyads occur naturally, such as in families or work settings, and where different individuals have different roles (e.g., father and daughter) or when interest centers on asymmetric relationships (e.g., supervisor and trainee), the exchangeability restrictions enforced by the model are no longer justified and we discuss how to relax them in Section \ref{sec:extdIRTc}.
A special case of non-exchangeability is where each dyad is composed of individuals from two different groups, such as husbands and wives,  and these groups are the same across dyads, so that there cannot, for example, be husband and wife dyads as well as father and daughter dyads. \citet{Kenny06} refer to this design as distinguishable dyads.

\begin{figure}[h]
\begin{center}
\includegraphics[width=0.50\textwidth,keepaspectratio]{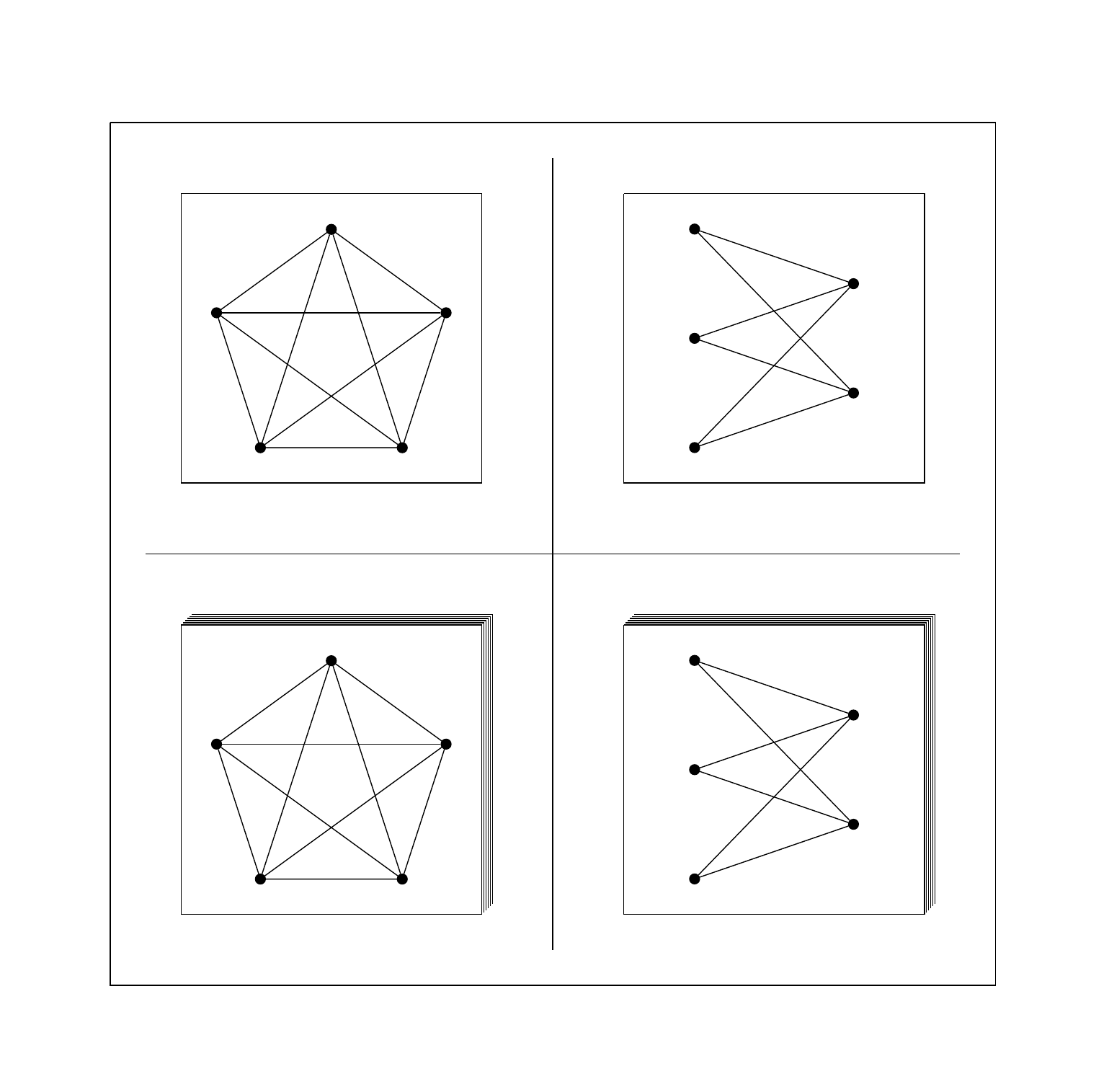}
\caption{Graphs representing round-robin design (upper-left panel), block design (upper-right panel), $k$-group round-robin design (lower-left panel), and $k$-group block design (lower-right panel). For the $k$-group designs each group is represented by a layer.}
\label{design}
\end{center}
\end{figure}

We now explore several dyadic designs for which the SRM is identified, following \citet{Kenny84} and \citet{Malloy86}.
The simplest and most common design is the round robin design.  In this design, each individual belongs to a dyad with every other member of the study, and there are a total of $\frac{n(n-1)}{2}$ dyads and $n(n-1)$ directed dyads.  In graph theoretic language where we view each individual as a node, the round-robin design is represented by a complete graph in the undirected case (see upper-left panel of Figure \ref{design}), and a complete directed graph (digraph) in the directed case.

One immediate extension of the round-robin design is the block design where the $n$ individuals are split into two blocks of sizes $p$ and $q$ respectively, and $p+q = n$.  Then, each individual from every block forms a dyad with every individual from the other block, but not with individuals in his/her block.  That is, there are a total of $pq$ undirected dyads and $2pq$ directed dyads.   In graph theoretic terms, such a design can be represented by a complete bipartite graph (see upper-right panel of Figure \ref{design}).  This occurs most naturally for distinguishable dyads, for example when interactions only occur between members of the opposite gender.  In this case, the $n$ individuals are split into two blocks by their gender.
\citet{Kenny06} refer to such a design as an asymmetric block design.

When individuals are nested in groups, such as families, work groups, or social networks, where each individual from the group forms a dyad with each other individual of the group, we have a  ``$k$-group round-robin design'' (see lower-left panel of Figure \ref{design}). In addition to such naturally occurring groups, the groups can also be created by the researcher to reduce response burden and costs by reducing the number of partners per actor and the number of dyads, respectively. Another reason for creating groups artificially is to allow individuals to interact within a group as a way to create the context for the dyadic responses. For example, \citet{christensen:98} created an initial social situation for groups of four lonely individuals that involved problem-solving tasks and subsequently collected dyadic ratings on personal characteristics. There can also be a block design within each group, resulting in the ``$k$-group block design'' (see lower-right panel of Figure \ref{design}).  This is the data design for the speed-dating application  in Section~\ref{sec:app}.

\subsection{Extended dIRT Model\label{sec:extdIRT}}
\subsubsection{Including Covariates for the Latent Traits\label{sec:extdIRTc}}
The dIRT model can be extended to take into account individual and dyadic covariates that may affect the latent traits at both the individual and dyadic levels by generalizing the idea of explanatory item response models \citep[e.g.,][]{DeBoeckWilson}.  One way that this can be accomplished is by specifying how  the means $\mu_\alpha$, $\mu_\beta$, and $\mu_\gamma$ depend on covariates, such as
\begin{align}
\mu_{\alpha,a} \ = \ \bm{x}_{\alpha,a}^{\prime}\bm{c}_\alpha,\quad \mu_{\beta,p} \ = \ \bm{x}_{\beta,p}^{\prime}\bm{c}_\beta,\quad \mu_{\gamma,a,p} \ = \ \bm{x}_{\gamma,a,p}^{\prime}\bm{c}_\gamma,
\end{align}
where $\bm{x}_{\alpha,a}$ are the covariates for $\alpha_a$,  $\bm{x}_{\beta,p}$ are the covariates for  $\beta_p$, 
$\bm{x}_{\gamma,a,p}$ are the covariates for $\gamma_{a,p}$, and $\bm{c}_\alpha$, $\bm{c}_\beta$ and $\bm{c}_\gamma$ are the corresponding regression coefficients. If the dyads are all pairs of individuals within a family ($k$-group round-robin design), the covariates can include dummy variables for the roles, e.g., for the actor being a father \citep{snijders:99}.

Keeping in mind that the response probability for actor $a$ when combined with partner $p$ is a function of the composite latent variable $\theta_{a,p}$, whose mean is $\mu_{\alpha,a}+ \mu_{\beta,p}+\mu_{\gamma,a,p}$, care must be taken to ensure that the regression coefficients are identified. For instance, if one of the covariates for $\mu_{\gamma,a,p}$ is the difference in some attribute, $z_a-z_p$, between the actor and partner, it is not possible to also include both the attribute for the actor, $z_a$, in the model for $\mu_{\alpha,a}$ and the attribute for the partner, $z_p$, in the model for $\mu_{\beta,p}$. Another example where identification is impossible is where dyads are
males paired only with females (i.e., if the actor is a male, then the partner must be a female and vice versa) and  gender is included as a covariate in the models for both $\mu_{\alpha,a}$ and $\mu_{\beta,p}$. Such an example is described in greater detail in Section \ref{sec:app}.

It is also possible to allow the variances of the latent traits to depend on covariates, for instance to have different variances for different roles within families \citep{snijders:99}. Such an approach allows modeling non-exchangeable dyads in general.

\subsubsection{Including Random Effects for the Latent Traits\label{sec:reffs}}
If the individuals are clustered in different ways, e.g., in schools and/or  neighborhoods, it may make sense to include cluster-level random effects into the models for $\alpha_a$ and $\beta_p$, to allow the actor and partner effects to be higher, on average, in some clusters than others, or, in other words, to have intraclass correlations. (For simplicity, we do not consider random effects in the model for $\gamma_{a,p}$.) 

An obvious specification would be to introduce corresponding cluster-level actor and partner effects, $A_j$ and $B_j$, respectively, for cluster $j$. The corresponding expression for $\theta_{a,p}$ then becomes
\[
\theta_{a,p}\ \equiv \ \alpha_a +  \beta_p + \gamma_{a,p}+A_{j[a]}+B_{j[p]},
\]
where $j[a]$ is the cluster that individual $a$ belong to. We could specify a bivariate normal distribution for $A_j,B_j$ with variances $\sigma_A^2$, $\sigma_B^2$ and correlation $\rho_{A,B}$.

These three additional parameters are identified if individuals in the same dyad can belong to different clusters. In this case,  $\cov(\theta_{a,p},\theta_{b,q})=0$ if the four different individuals $a$, $p$, $b$, $q$  come from four different clusters.
Otherwise, we add $\sigma_A^2$ to the covariance if and only if (iff) $j[a]=j[b]$, $\sigma_B^2$ iff $j[p]=j[q]$,  $\rho_{A,B}\sigma_A\sigma_B$ iff either $j[a]=j[q]$ or $j[p]=j[b]$ but not both, and $2\rho_{A,B}\sigma_A\sigma_B$ iff
$j[a]=j[q]$ and $j[p]=j[b]$. Depending on the cluster memberships of these four individuals, each of these terms can be added in isolation or in combination, producing eight distinct covariances. The  parameters $\sigma_A^2$, $\sigma_B^2$ and $\rho_{A,B}$ are identified from these reduced form parameters alone. 
Further distinct covariances arise if, for instance, the actor is the same individual in both dyads. In this case,
$\cov(\theta_{a,p},\theta_{a,q})=\sigma_a^2+\sigma_A^2$ if the different individuals, $a$, $p$ and $q$, all belong to different clusters and we follow the same rules as above for adding the other terms besides $\sigma_A^2$. 

However, if dyads are formed only among individuals within the same cluster, e.g., students are paired only with other students from the same school, then the term $\sigma_A^2+\sigma_B^2+2\rho_{A,B}\sigma_A\sigma_B$ appears in all variances and covariances unless the two dyads belong to different clusters. This can occur only if the two dyads do not share any individuals in common, in which case we obtain  $\cov(\theta_{a,p},\theta_{b,q})=\sigma_A^2+\sigma_B^2+2\rho_{A,B}\sigma_A\sigma_B$ if dyad ($a$, $p$) belongs to the same cluster as dyad ($b$, $q$)  and $\cov(\theta_{a,p},\theta_{b,q})=0$, otherwise. It follows that only the sum $\sigma_A^2+\sigma_B^2+2\rho_{A,B}\sigma_A\sigma_B$
is identified and therefore it makes sense to define  $u_j \equiv A_j+B_j$, with one variance parameter $\sigma_u^2$, and to include $u_j$ directly in the model for $\theta_{a,p}$.

It is of course possible to handle multiple nested or non-nested classifications by adding the corresponding random intercepts $u$ if dyads are formed within a classification and $A$ and $B$ if dyads are formed across classifications (e.g., neighborhood when dyads are formed within schools or firms). Non-exchangeability can be handled by specifying different (co)variances for  $u$ or for $A$ and $B$ for different groups of individuals.

\subsubsection{Distal Outcomes}
The dIRT model can be extended by using, for instance, Generalized Linear Models to model one or more distal outcomes, where $\alpha_a$, $\beta_p$, and $\gamma_{a,p}$ are latent covariates.  

For example, we can consider a binary distal outcome $d_{a, p}$ of a dyad $(a, p)$ taking the value of 1 with the conditional probability $\pi_{a, p}$ given the latent traits, and 0 otherwise.  For the speed-dating application considered in Section 4, the distal outcome is whether each actor in a dyad elected to see the partner again. Here, $\pi_{a, p}$ can be modeled using the logistic regression
\begin{align}
\label{distal}
\log \left( \frac{\pi_{a, p}}{1-\pi_{a, p}}\right) \ = \  b_0 &+ b_1\alpha_a + b_2\alpha_p + b_3\beta_a + b_4\beta_p + b_5\gamma_{a, p} + b_6\gamma_{p, a} \nonumber\\[-3mm] 
&+b_7\alpha_a\alpha_p + b_8\beta_a\beta_p + b_9\gamma_{a, p}\gamma_{p, a}.
\end{align}
Distal outcome regressions can also include covariates of both the individual and the dyad if necessary.

Notice that in the above example, the distal outcome is directed in the sense that it depends on which individual in the dyad plays the role of the actor, and the individuals are therefore not exchangeable in the sense that the effect of $\alpha_a$ on the distal outcome for actor $a$ is not necessarily the same as the effect of $\alpha_p$. If the distal outcome is undirected, however, and individuals within dyads are exchangeable (e.g., in the case of pairs of individuals participating in a collaborative problem solving task where the outcome of interest is how well the task was completed per pair), then, (\ref{distal}) should be constrained to have $b_1 = b_2$, $b_3 = b_4$, and $b_5 = b_6$. If there is one undirected outcome per dyad and the individuals in the dyad are non-exchangeable (e.g., males paired with females), such a constraint is not needed if, for instance, $a$ represents the male and $p$ the female in the dyad.

\subsection{Relationship with Other Models \label{sec:other}}

We first review models for dyadic designs for which the dIRT and SRM are not identified, either because individuals can belong only to one dyad or because actor/partner role reversal is not possible.

Starting with the situation where individuals belong to only one grouping (dyad or larger group), the dIRT model reduces to a multilevel IRT where $\theta_{a, g} = \zeta_g + \zeta_{a, g}$, sometimes called a variance components factor/IRT model \citep[e.g.,][]{Anders2004}. Here $\zeta_g$ is a group-level random intercept and $\zeta_{a, g}$ an individual-level random
 intercept.
 
In the dyadic data literature, the most popular model for this case is the Actor-Partner-Interdependence Model (APIM) proposed by \citet{kenny:96}. The APIM for distinguishable dyads is basically a bivariate regression model where the actor's and partner's continuous responses $y_a$ and $y_p$ are both regressed on the covariates $x_a$ and $x_p$ of both the actor and the partner: 
\begin{displaymath}
    y_{a}  \ = \   b_1 x_{a} + c_1 x_{p} + \zeta_{a}, \ \ \
    y_{p} \ = \   c_2 x_{a} + b_2 x_{p} + \zeta_{p},
\end{displaymath}
where the disturbances $\zeta_{a}$ and $\zeta_{p}$ are correlated. Here, $b_1$ and $b_2$ are interpreted as actor effects and $c_1$ and $c_2$ as partner effects. In the exchangeable APIM the actor effects are constrained to be equal, $b_1=b_2$, as are the partner effects, $c_1=c_2$, and the variances, $\Var(\zeta_{a})=\Var(\zeta_{p})$. 
Generalizations of the classical APIM have also been proposed.  For example, \citet{loeys:2013} used generalized linear mixed models for categorical $y_a$ and $y_p$. 
\citet{Alex} replaced the observed variables, $x_a$, $x_p$ and $y_a$, $y_p$, in the APIM  by latent variables measured by multiple items via IRT models.

The mutual-influence model \citep{kenny:96} has no partner covariate effects which allows a reciprocal or mutual relationship between the responses of the actor and partner:
\begin{displaymath}
    y_{a} \ = \   d_1 y_{p} + b_1 x_{a}  + \zeta_{a}, \ \ \
    y_{p}  \ = \   d_2 y_{a}  + b_2 x_{p} + \zeta_{p},
\end{displaymath}
where $\zeta_{a}$ and $\zeta_{p}$ are correlated. 
This is a simultaneous equation model where $d_1$ and $d_2$
represent the mutual influence between the responses in a pair
and $x_{a}$ and $x_{p}$ serve as instrumental variables for the endogenous explanatory variables $y_{a}$ and $y_{p}$, respectively.
In the exchangeable version \citep{duncan:68}, the actor effects are constrained to be equal, $b_1=b_2$, as are the mutual effects, $d_1=d_2$, and the variances, $\Var(\zeta_{a})=\Var(\zeta_{p})$.

In the Common-Fate Model (CFM) of \citet{Kenny84} a dyad-level latent variable $\eta_g$ for dyad $g$, measured by the continuous responses $y_{a,g}$ and $y_{p,g}$, is regressed on a dyad-level latent variable $\xi_g$, measured by the continuous covariates $x_{a,g}$ and $x_{p,g}$:
\begin{displaymath}
    x_{a,g}  \  = \   \xi_g + \delta_{a,g}, \ \ \  x_{p,g}  \ = \  \xi_g + \delta_{p,g}, \ \ \
    y_{a,g} \  = \  \eta_g + \epsilon_{a,g}, \ \ \ y_{p,g} \  = \  \eta_g + \epsilon_{p,g},
\end{displaymath}
\begin{displaymath}
    \eta_g    \ = \   \gamma\xi_g + \zeta_g.
\end{displaymath}
The unique factors $\delta_{a,g}$ and $\epsilon_{a,g}$ for the actor-variables are correlated as are   $\delta_{p,g}$ and $\epsilon_{p,g}$ for the partner-variables. Hence, the relationships between the variables is decomposed into a dyad-level relation (represented by $\gamma$) and two individual-level relations (represented by the error covariances, $\cov(\delta_{a,g},\epsilon_{a,g})$ and $\cov(\delta_{p,g},\epsilon_{p,g})$). In the exchangeable case, the following constraints are necessary: $\Var(\delta_{a,g})=\Var(\delta_{p,g})$,  $\Var(\epsilon_{a,g})=\Var(\epsilon_{p,g})$, $\cov(\delta_{a,g},\epsilon_{a,g})=\cov(\delta_{p,g},\epsilon_{p,g})$.
To use the CFM in an IRT framework, \cite{Alex} simply allowed all items measuring the latent versions of $x_{a,g}$
and $x_{p,g}$  to load on $\xi_g$ and all items measuring the latent versions of $y_{a,g}$ and $y_{p,g}$  to load on $\eta_g$. We believe that a more appropriate approach would have been to replace each of $x_{a,g}$, $x_{p,g}$, $y_{a,g}$ and $y_{p,g}$ by a separate (first-order) latent variable, so that $\xi_g$ and $\eta_g$ become second-order latent variables and the error covariances of the CFM can be directly accommodated as covariances among the disturbances of the first-order latent variables.

We now discuss the situation where individuals appear in multiple dyads but actor/partner role reversals (or reciprocals) do not occur. For example,  if the dyads are raters and examinees (with each examinee rated by several raters and each rater rating several examinees) only the raters  provide responses
so that the raters are always the actors and the examinees are always the partners. Then $\alpha_a$ is the rater leniency, $\beta_p$ the examinee ability and  $\gamma_{a,p}$,  interpretable as person-specific rater leniency, can be included only if  raters assesses several items by the same examinee (see, e.g., \citeauthor{shin:2019}, 2019). In such a design, $\rho_{\alpha\beta}$ and $\rho_{\gamma}$ are not defined because examinees and raters never switch roles.  

We now turn to designs of the kind discussed in Section~\ref{sec:design},
where the SRM or dIRT are identified.  \citet{Kenny84} defined the original SRM for a continuous observed outcome $y$ of actor $a$ in the presence of partner $p$ measured over multiple time points $t$ as 
\begin{equation*}
    y_{a,p,t} \ = \ \alpha_{a} + \beta_{p} + \gamma_{a, p} + \epsilon_{a, p, t}.
\end{equation*}
Here, $\epsilon_{a, p, t}$ can be viewed as test-retest measurement error. Since this term reduces to $\epsilon_{a, p}$ if there is only one time point, the identifiability of the above model, and in particular the variance of $\gamma_{a, p}$ separately from that of $\epsilon_{a,p,i}$, hinges crucially on measurements of the same dyad across multiple time points. In the dIRT model, multiple items essentially play the role of multiple time points, allowing for identification of the variance of $\gamma_{a, p}$.  If one does not have multiple items or time-points, the model may still be identifiable if we assume that for two individuals $a$ and $p$, $\gamma_{a, p} = \gamma_{p, a}$.  That is, we assume that the dyadic effect of the pair is symmetric across the role that the individuals play and in that sense is no longer directional. In this case, $\gamma_{a,p}$ simply induces additional dependence between the responses for a given dyad and we can alternatively replace $\gamma_{a, p} + \epsilon_{a, p}$ by a single error term, typically denoted  $\gamma_{a, p}$, that is correlated across members of the same dyad. In a $k$-group round-robin design, it may also make sense to include a group-level random intercept, for instance, when the groups are families, with each pair of family members forming a dyad \citep{snijders:99,loncke:2018}. Such a model is described at the end of Section~\ref{sec:reffs}.

In genetic experiments, a diallel cross is the set of all possible matings between several genotypes. The genotypes may be defined as individuals, clones, homozygous lines, etc. \citep{hayman:54}.
Some quantitative trait is measured for offspring from father $a$ and mother $p$, and there are reciprocal crosses, with the role of mother and father reversed. Li and Loken~(\citeyear{li:2002}) show the correspondence between the SRM and a diallel model used in genetics  \citep[e.g.,][]{cockerham:77}:
\[
  y_{a,p} \ = \ \mu+g_{a}+g_{p}+s_{a,p}+d_a-d_p+r_{a,p},
\]
where all terms are uncorrelated, except for $g_a$ and $d_a$, and where $s_{a,p}=s_{p,a}$
and $r_{a,p}=-r_{p,a}$.  The correspondence with the SRM is that  $\alpha_a=g_a+d_a$, $\beta_p=g_p-d_p$, and $\gamma_{a,p}=s_{a,p}+r_{a,p}$.

Multivariate extensions of the SRM have been proposed for the situation where actors provide ratings on several continuous variables \citep{nestler:2018,luedtke:2018}. For the case with a single time-point, the model can be written as
\begin{equation*}
    y_{a,p,i} \ = \ \alpha_{a,i} + \beta_{p,i} + \gamma_{a, p,i}
\end{equation*}
for variable $i$, where $\epsilon_{a,p,i}$ has been removed because only one error term (correlated across members of the same dyad) can be included. Unstructured covariance matrices are specified for each of these terms across variables and, in addition to the same-variable covariances between $\alpha_{a,i}$ and $\beta_{a,i}$ and between $\gamma_{a,p,i}$ and $\gamma_{p,a,i}$ that are part of a univariate SRM, the model allows for all corresponding cross-variable covariances as well. As far as we know, the common factor analogue to our dIRT model (where the measurement model for $\theta_{a,p}$ is a univariate factor model) has not been discussed in the literature.

We are aware of only very few papers that extend the classic SRM model to handle non-continuous responses, such as \citet{koster:2014} who used bivariate Poisson models for counts and \citet{koster:2018} who used bivariate probit models for binary responses.

\section{Estimation\label{sec:est}}

The dIRT model includes crossed random effects so that the marginal likelihood involves high-dimensional integrals. For example, in a $k$-group block design, the dimensionality of integration for the likelihood contribution of a group
is $p+1$ or $q+1$, whichever is smaller~\citep{goldstein:87}. Numerical integration or Monte Carlo 
integration quickly becomes
prohibitive and approximate methods are often not satisfactory (see, e.g., \citeauthor{jeon:2017}, 2017 and references therein). Fortunately, Bayesian estimation via  Markov-chain Monte Carlo (MCMC) is feasible, and we adopt this approach here.
Specifically, we use the the ``No-U-Turn'' sampler \citep{NUTS} 
 implemented in \texttt{Stan} \citep{Stan}.  The \texttt{Stan} language affords us great flexibility in extending the basic dIRT model.    We also verified all results using \verb|Matlab| (version r2016b) via custom-written code based on the Metropolis-Hastings algorithm \citep{MH}.

To use MCMC, we define prior distributions for the parameters in (\ref{hyperparameters}) as well as the item parameters in (\ref{dirt}) (and potentially the coefficients of the distal outcome regression in (\ref{distal})). In our approach, we take the distributions of all hyperparameters $\sigma_\alpha^2, \sigma_\beta^2, \sigma_\gamma^2, \rho_{\alpha\beta}$ and $\rho_\gamma$ to be noninformative by assuming uniform distributions for the variances $[0, +\infty)$ and correlations $[-1, 1]$. For step difficulties $\delta_{i, j}$ and regression coefficients $b_0, b_1, \ldots, b_9$ in the distal outcome model (\ref{distal}), we specify noninformative uniform priors $(-\infty , +\infty)$.

All parameter estimates were obtained using MCMC simulations of 4 chains with $2,000$ iterations, with a burn-in period of $1,000$ iterations.  The parameter and hyper-parameter estimates are expected a posteriori (EAP) values obtained as means over the converged (post burn-in) MCMC draws for the four chains, i.e., they are based on an MCMC sample size of 4,000.  Convergence was assessed by monitoring the $\hat{R}$ statistic \citep{GelmanRubin1992}.

\label{joint} The distal outcome model in (\ref{distal}) can be estimated  jointly with the dIRT model by combining the log-likelihood contributions from the dIRT ($\ell_{\textnormal{dIRT}}$) and distal outcome ($\ell_{\textnormal{distal}}$) models in forming the joint log posterior of all parameters, given the dIRT item responses and distal outcome. 

Joint estimation of the dIRT  and distal outcome  models is consistent and asymptoticaly efficient if both models are correctly specified.  However, to protect against misspecification of the distal outcome (or ``structural'') model, a sequential approach could be used where the parameters of the dIRT (``measurement'') model are estimated in step 1 and subsequent steps are used to obtain estimates of the structural (distal outcome) model parameters. If the measurement model is correctly specified, the estimates from step 1 are consistent even if the structural model is misspecified. However, if the structural model is correctly specified, joint estimation is more efficient than sequential approaches. 
From a conceptual point of view, it has been argued in the structural equation modeling and latent class literature, that altering the structural model by, for instance, adding or removing distal outcomes, affects the interpretation of the measurement model because these distal outcomes play a similar role to the items or indicators that define the latent traits. Sequential modeling can protect against such ``interpretational confounding'' \citep{Burt} where the meaning of a construct is different from the meaning intended by the researcher (see \citet{KuhaBakk} for further discussion). 

The most obvious sequential approach is to use factor score regression \citep{skrondal:2001} where one estimates the measurement model (step 1), obtains judiciously chosen scores for the latent traits from the measurement model (step 2), and substitutes these scores for the latent traits to estimate the structural model as if the latent traits were observed (step 3). This approach was adopted by \citet{loncke:2018} for SRMs. However, factor score regression is only consistent for link functions that are rarely of relevance in IRT (such as the identity) and naive standard errors from this approach are moreover underestimated. To address these limitations, a multiple imputation approach can be used, where multiple draws of the latent traits are obtained from their posterior distribution and the estimates
for the structural model are combined using Rubin's formula \citep{rubin}. \citet{luedtke:2018} use such an approach in an SRM to estimate covariate effects on individual-level latent traits (i.e., as discussed in Section \ref{sec:extdIRTc}). Multiple imputation is natural in a Bayesian setting where full posteriors of the latent traits are available. A more straightforward pseudo-likelihood estimator, in the sense of \citet{gong:81}, was proposed by  \citet{SkrondalKuha} (see also \citet{KuhaBakk}). In this case the measurement model is first estimated, followed by joint estimation of the measurement and structural models under the constraint that the parameters of the measurement model are set equal to the estimates from the first stage. 

 We present the results of the joint approach in this paper and include results for the sequential approach with multiple imputation in Appendix B.

\section{Speed-Dating Application \label{sec:app}}

We use a speed-dating dataset \citep{Fisman} to examine the mutual attractiveness ratings of both individuals in a dyad to look for evidence of interactions that cannot be explained solely by the individuals' attractiveness or rating preferences.  We also considered whether males and females differ in how they perceive their interactions. Additionally, by treating the final dating decision of whether the actor wants to see the partner again as a distal outcome, we investigate to what extent it relates to the dyadic latent trait.

The data was collected at 21 separate researcher-organized speed-dating sessions, over a period of 2 years, with 10-44 students from graduate and professional schools at Columbia University in each session. During these sessions, attended by nearly an equal number of male and female participants, all members of one gender would meet and interact with every member of the opposite gender for 5 minutes each.  At the end of the 5 minute session, participants would rate their partner based on five attractiveness factors on a form attached to a clipboard that they were provided with.  They also indicated whether or not they would like to see the person again.

After data cleaning, we had a total of 551 individuals, interacting in 4,184 distinct pairs, leading to 8,368 surveys completed (twice the number of pairs, given that both members of a pair rated each other). This corresponds to the ``$k$-group block-dyadic'' design described in Section \ref{sec:design}.  An illustrative example of data collected for one item in a balanced group of 10 individuals is shown in Figure \ref{fig:checker}. 

\begin{figure}[h]
\begin{center}
\begin{tabular}{cccccccccccc}
&          & \multicolumn{10}{c}{partner, $p$}\\
&          & $F_1$       & $F_2$       & $F_3$       & $F_4$       & $F_5$       & $M_6$      & $M_7$      & $M_8$      & $M_9$      & $M_{10}$    \\
\cline{3-12}
\multirow{10}{*}{actor, $a$}
& $F_1$    & \multicolumn{1}{|c|}{\cellcolor{gray}}            & \multicolumn{1}{|c|}{\cellcolor{gray}}             & \multicolumn{1}{|c|}{\cellcolor{gray}}                & \multicolumn{1}{|c|}{\cellcolor{gray}}            & \multicolumn{1}{|c|}{\cellcolor{gray}}             & \multicolumn{1}{|c|}{$y_{1, 6}$} 
           & \multicolumn{1}{|c|}{$y_{1, 7}$}                   & \multicolumn{1}{|c|}{$y_{1, 8}$}                    & \multicolumn{1}{|c|}{$y_{1, 9}$} 
           & \multicolumn{1}{|c|}{$y_{1, 10}$} \\
\cline{3-12}
& $F_2$    & \multicolumn{1}{|c|}{\cellcolor{gray}}            & \multicolumn{1}{|c|}{\cellcolor{gray}}             & \multicolumn{1}{|c|}{\cellcolor{gray}}                & \multicolumn{1}{|c|}{\cellcolor{gray}}            & \multicolumn{1}{|c|}{\cellcolor{gray}}             & \multicolumn{1}{|c|}{$y_{2, 6}$} 
           & \multicolumn{1}{|c|}{$y_{2, 7}$}                   & \multicolumn{1}{|c|}{$y_{2, 8}$}                    & \multicolumn{1}{|c|}{$y_{2, 9}$} 
           & \multicolumn{1}{|c|}{$y_{2, 10}$} \\ 
\cline{3-12}
& $F_3$    & \multicolumn{1}{|c|}{\cellcolor{gray}}            & \multicolumn{1}{|c|}{\cellcolor{gray}}             & \multicolumn{1}{|c|}{\cellcolor{gray}}                & \multicolumn{1}{|c|}{\cellcolor{gray}}            & \multicolumn{1}{|c|}{\cellcolor{gray}}             & \multicolumn{1}{|c|}{$y_{3, 6}$} 
           & \multicolumn{1}{|c|}{$y_{3, 7}$}                   & \multicolumn{1}{|c|}{$y_{3, 8}$}                    & \multicolumn{1}{|c|}{$y_{3, 9}$} 
           & \multicolumn{1}{|c|}{$y_{3, 10}$} \\ 
\cline{3-12}
& $F_4$    & \multicolumn{1}{|c|}{\cellcolor{gray}}            & \multicolumn{1}{|c|}{\cellcolor{gray}}             & \multicolumn{1}{|c|}{\cellcolor{gray}}                & \multicolumn{1}{|c|}{\cellcolor{gray}}            & \multicolumn{1}{|c|}{\cellcolor{gray}}             & \multicolumn{1}{|c|}{$y_{4, 6}$} 
           & \multicolumn{1}{|c|}{$y_{4, 7}$}                   & \multicolumn{1}{|c|}{$y_{4, 8}$}                    & \multicolumn{1}{|c|}{$y_{4, 9}$} 
           & \multicolumn{1}{|c|}{$y_{4, 10}$} \\  
\cline{3-12}
& $F_5$    & \multicolumn{1}{|c|}{\cellcolor{gray}}            & \multicolumn{1}{|c|}{\cellcolor{gray}}             & \multicolumn{1}{|c|}{\cellcolor{gray}}                & \multicolumn{1}{|c|}{\cellcolor{gray}}            & \multicolumn{1}{|c|}{\cellcolor{gray}}             & \multicolumn{1}{|c|}{$y_{5, 6}$} 
           & \multicolumn{1}{|c|}{$y_{5, 7}$}                   & \multicolumn{1}{|c|}{$y_{5, 8}$}                    & \multicolumn{1}{|c|}{$y_{5, 9}$} 
           & \multicolumn{1}{|c|}{$y_{5, 10}$} \\ 
\cline{3-12}
& $M_6$    & \multicolumn{1}{|c|}{$y_{6, 1}$}                   & \multicolumn{1}{|c|}{$y_{6, 2}$}                    & \multicolumn{1}{|c|}{$y_{6, 3}$}                      & \multicolumn{1}{|c|}{$y_{6, 4}$}                   & \multicolumn{1}{|c|}{$y_{6, 5}$}                    & \multicolumn{1}{|c|}{\cellcolor{gray}}                & \multicolumn{1}{|c|}{\cellcolor{gray}}            & \multicolumn{1}{|c|}{\cellcolor{gray}}             & \multicolumn{1}{|c|}{\cellcolor{gray}}                & \multicolumn{1}{|c|}{\cellcolor{gray}} \\
\cline{3-12}
& $M_7$    & \multicolumn{1}{|c|}{$y_{7, 1}$}                   & \multicolumn{1}{|c|}{$y_{7, 2}$}                    & \multicolumn{1}{|c|}{$y_{7, 3}$}                      & \multicolumn{1}{|c|}{$y_{7, 4}$}                   & \multicolumn{1}{|c|}{$y_{7, 5}$}                    & \multicolumn{1}{|c|}{\cellcolor{gray}}                & \multicolumn{1}{|c|}{\cellcolor{gray}}            & \multicolumn{1}{|c|}{\cellcolor{gray}}             & \multicolumn{1}{|c|}{\cellcolor{gray}}                & \multicolumn{1}{|c|}{\cellcolor{gray}} \\
\cline{3-12}
& $M_8$    & \multicolumn{1}{|c|}{$y_{8, 1}$}                   & \multicolumn{1}{|c|}{$y_{8, 2}$}                    & \multicolumn{1}{|c|}{$y_{8, 3}$}                      & \multicolumn{1}{|c|}{$y_{8, 4}$}                   & \multicolumn{1}{|c|}{$y_{8, 5}$}                    & \multicolumn{1}{|c|}{\cellcolor{gray}}                & \multicolumn{1}{|c|}{\cellcolor{gray}}            & \multicolumn{1}{|c|}{\cellcolor{gray}}             & \multicolumn{1}{|c|}{\cellcolor{gray}}                & \multicolumn{1}{|c|}{\cellcolor{gray}} \\
\cline{3-12}
& $M_9$    & \multicolumn{1}{|c|}{$y_{9, 1}$}                   & \multicolumn{1}{|c|}{$y_{9, 2}$}                    & \multicolumn{1}{|c|}{$y_{9, 3}$}                      & \multicolumn{1}{|c|}{$y_{9, 4}$}                   & \multicolumn{1}{|c|}{$y_{9, 5}$}                    & \multicolumn{1}{|c|}{\cellcolor{gray}}                & \multicolumn{1}{|c|}{\cellcolor{gray}}            & \multicolumn{1}{|c|}{\cellcolor{gray}}             & \multicolumn{1}{|c|}{\cellcolor{gray}}                & \multicolumn{1}{|c|}{\cellcolor{gray}} \\
\cline{3-12}
& $M_{10}$ & \multicolumn{1}{|c|}{$y_{10, 1}$}                  & \multicolumn{1}{|c|}{$y_{10, 2}$}                   & \multicolumn{1}{|c|}{$y_{10, 3}$}                      & \multicolumn{1}{|c|}{$y_{10, 4}$}                  & \multicolumn{1}{|c|}{$y_{10, 5}$}                   & \multicolumn{1}{|c|}{\cellcolor{gray}}                & \multicolumn{1}{|c|}{\cellcolor{gray}}            & \multicolumn{1}{|c|}{\cellcolor{gray}}             & \multicolumn{1}{|c|}{\cellcolor{gray}}                & \multicolumn{1}{|c|}{\cellcolor{gray}} \\
\cline{3-12}
\end{tabular}
\caption{Example of responses $y_{a, p}$ of actor $a$ rating partner $p$ in a single-group block-dyadic structure consisting of 5 females $F_1, \ldots, F_5$ and 5 males $M_6, \ldots, M_{10}$.}
\label{fig:checker}
\end{center}
\end{figure}

In the data, the rating by actor $a$ of partner $p$ on item $i$ is given by $y_{a, p, i}$. Each item was rated on a 10-point Likert-scale, which we collapsed to a 5-point scale by combining pairs of adjacent response categories to mitigate sparseness.  Participants rated each other on 5 different items, all related to the overall attractiveness of the partner (viz. physical attractiveness, ambition, how fun they were, intelligence, and sincerity).  We dropped all invalid ratings from an actor of a partner and the corresponding ratings from the partner of the actor even if the latter was valid.  This amounted to a loss of less than 5\% of the data.

In addition to each individual's rating of his/her partner, we also had access to an indicator $d_{a,p}$ for whether actor $a$ elected to see partner $p$ again.  Note that this indicator is directional and $d_{a,p}$ may therefore differ from $d_{p, a}$.  However, embedding the dIRT model within a distal outcome regression where the distal outcome is non-directional is also possible.  For example, if we knew whether the dyad did in fact go on a date, this outcome would be unique to the dyad.

\label{speeddatingemp}
Using the joint MCMC estimation approach described in Section \ref{joint}, the results of estimating the basic dyadic partial credit model (\ref{dirt}) and the model also including a distal regression (\ref{distal}) are presented in Tables \ref{varcorest} and \ref{distalest} under the heading ``without gender''.  
The code used to obtain the subsequent results is provided in Appendix A and explained in a \texttt{Stan} case-study \citep{stancasestudy}.
We estimate two versions of the distal regression, one with all 10 parameters $b_0, \ldots, b_9$ (labeled ``with interactions''), and another model with $b_7=b_8=b_9=0$ (labeled ``without interactions").    The estimates presented are the posterior means of the MCMC draws, and the values in parentheses represent the $2.5^{\text{th}}$ and the $97.5^{\text{th}}$ quantiles of the posterior distribution of the MCMC draws. 
\begin{table}[h]
\caption{Estimates of Standard Deviations and Correlations of Individual and Dyadic Latent Traits (Joint Approach)}
\label{varcorest}
\begin{center}
\begin{tabular}{|l|.,|.,|.,|}
\hline
\multirow{ 2}{*}{}   & \multicolumn{4}{|c|}{without gender}      & \multicolumn{2}{|c|}{with gender}\\ \cline{2-7}
                     & \multicolumn{2}{|c|}{with interactions}   & \multicolumn{2}{|c|}{without interactions}         & \multicolumn{2}{|c|}{without interactions}\\
\hline
$\mu_{\text{male}}$  &         &                 &         &                & $ 0.08$ & $(-0.10,0.24)$\\
$\sigma_\alpha$      & $ 1.03$ & $( 0.96,1.10)$  & $ 1.03$ & $( 0.96,1.10)$ & $ 1.03$ & $( 0.96,1.10)$\\
$\sigma_\beta$       & $ 0.63$ & $( 0.58,0.68)$  & $ 0.63$ & $( 0.58,0.68)$ & $ 0.63$ & $( 0.58,0.69)$\\
$\sigma_\gamma$      & $ 0.98$ & $( 0.95,1.02)$  & $ 0.98$ & $( 0.95,1.01)$ & $ 0.98$ & $( 0.95,1.02)$\\
$\rho_{\alpha\beta}$ & $-0.06$ & $(-0.17,0.04)$  & $-0.06$ & $(-0.16,0.04)$ & $-0.07$ & $(-0.17,0.03)$\\
$\rho_\gamma$        & $ 0.46$ & $( 0.42,0.51)$  & $ 0.46$ & $( 0.41,0.51)$ & $ 0.46$ & $( 0.42,0.51)$\\
\hline
\end{tabular}
\end{center}
\end{table}

\begin{table}[h]
\caption{Estimates for Distal Outcome Regression (Joint Approach)}
\label{distalest}
\begin{center}
\begin{tabular}{|l|.,|.,|.,|}
\hline
\multirow{ 2}{*}{}   & \multicolumn{4}{|c|}{without gender}      & \multicolumn{2}{|c|}{with gender}\\ \cline{2-7}
                     & \multicolumn{2}{|c|}{with interactions}   & \multicolumn{2}{|c|}{without interactions}         & \multicolumn{2}{|c|}{without interactions}\\
\hline
$b_0$ & $-0.87$ & $(-1.03,-0.71)$ & $-0.88$ & $(-1.04,-0.73)$ & $-0.88$ & $(-1.04,-0.73)$ \\
$b_1$ & $ 0.15$ & $(-0.04, 0.33)$ & $ 0.14$ & $(-0.05, 0.32)$ & $ 0.14$ & $(-0.04, 0.32)$ \\
$b_2$ & $-0.02$ & $(-0.13, 0.09)$ & $-0.02$ & $(-0.13, 0.09)$ & $-0.03$ & $(-0.13, 0.08)$ \\
$b_3$ & $-3.03$ & $(-3.62,-2.58)$ & $-2.92$ & $(-3.46,-2.49)$ & $-2.91$ & $(-3.44,-2.45)$ \\
$b_4$ & $ 3.56$ & $( 3.17, 4.01)$ & $ 3.48$ & $( 3.12, 3.93)$ & $ 3.48$ & $( 3.11, 3.91)$ \\
$b_5$ & $ 3.50$ & $( 3.06, 4.06)$ & $ 3.42$ & $( 3.00, 3.95)$ & $ 3.42$ & $( 2.99, 3.94)$ \\
$b_6$ & $ 0.17$ & $( 0.00, 0.35)$ & $ 0.13$ & $(-0.04, 0.29)$ & $ 0.13$ & $(-0.04, 0.29)$ \\
$b_7$ & $-0.01$ & $(-0.13, 0.09)$ & $     $ & $             $ & $     $ & $             $ \\
$b_8$ & $ 0.45$ & $( 0.06, 0.87)$ & $     $ & $             $ & $     $ & $             $ \\
$b_9$ & $-0.28$ & $(-0.53,-0.02)$ & $     $ & $             $ & $     $ & $             $ \\
\hline
\end{tabular}
\end{center}
\end{table}

\subsection{Partitioning of Variance between Individual and Dyadic Latent Traits}
Standard deviation and correlation estimates are reported in Table \ref{varcorest}. In the dIRT, the variance of the composite latent variable $\theta_{a,p}$ is the sum of the variances of the individual and dyad-level latent traits, $\alpha_a$, $\beta_p$ and $\gamma_{a, p}$. It is instructive to examine the relative contributions of these latent traits to the composite.  
The percentage of the variance of $\theta_{a,p}$ that is due to  $\alpha_a$, $\beta_p$ and $\gamma_{a, p}$ is estimated as 44\%, 16\% and 40\%, respectively.

Interestingly, the variance of $\alpha_a$ is larger than that of $\beta_p$, implying that the actor's perception of the partner is more influenced by the actor's average tendency to rate others as attractive, which we could call actor leniency, than by the partner's average tendency to be rated as attractive, which we could think of as the partner's ``universal'' attractiveness.    While the majority (60\%) of the variance is accounted for by the individual effects ($\alpha_a$ and $\beta_p$), the dyadic effect $(\gamma_{a, p})$ accounts for a substantial proportion of the total variance, at 40\%.  A traditional IRT model, measuring individual latent traits only, would ignore this contribution, which can be thought of as the ``eye-of-the-beholder'' effect.  In particular, this dyadic component would not be identifiable for standard IRT data where the individual only belongs to a single dyad.

\subsection{Correlations}
The within-person correlation $\rho_{\alpha\beta}$ of $\alpha_a$ and $\beta_a$ reflects the relationship between how willing an individual was to rate someone else as attractive (``leniency''), and his/her own attractiveness. If this correlation is positive, it indicates that the more attractive an individual is, the more lenient he/she is in his/her ratings. If negative, it indicates that more attractive an individual is, the harsher he/she tends to be in rating his/her partners' attractiveness.

The between-person correlation $\rho_\gamma$ of a dyad reflects the extent to which the (directed) dyadic trait is correlated between members of a given pair. If positive, it indicates that when an individual is affected by a social interaction with his/her partner, the partner will be more likely to also be affected in a similar manner. If negative, it suggests that members of a pair perceive their interaction in opposing ways.

Table \ref{varcorest} shows that the estimate of the correlation $\rho_\gamma$ is positive with a 95\% credible interval that does not contain zero.  In contrast, the  estimate of the correlation $\rho_{\alpha\beta}$ is negative with a 95\% credible interval containing zero.  The relatively larger estimated between-individual correlation indicates that members of each pair were likely to perceive their interaction similarly.     

\subsection{Distal Outcome Regression}
We estimate the distal outcome regression for each individual's dating decision in (\ref{distal}) using the joint approach described in Section \ref{joint} and compare the regression estimates in Table \ref{distalest} for the full model (under ``with interactions'') and a reduced model without interaction terms (under ``without interactions'').

We see that the estimated distal outcome regression coefficients are largest, in absolute value, for: a) the individual attractiveness $\alpha$ of both the actor ($\hat{b}_3$) and the partner ($\hat{b}_4$), and b) the unique relationship of the dyad $\gamma$ from the actor's perspective ($\hat{b}_5$) but {\em not} for that from the partner's perspective  ($\hat{b}_6$). Finding b) is consistent with our expectations given that the distal outcome reflects the viewpoint of the actor, rather than that of the partner. However, a less obvious finding is a) because the rater's own attractiveness, $\hat{b}_3$, {\em negatively} influences their dating decision. This suggests that the more attractive a rater was, the less likely they were to want to see the partner again. The estimated coefficients are tiny for the leniency of both the actor and the partner, as well as for the unique relationship of the dyad from the perspective of the partner, and have 95\% credible intervals either containing zero or having one limit close to zero.

\subsection{Gender Differences}
Both the basic dIRT model and the model with a distal outcome can be extended to account for differences in the way females and males perceived their social interactions. In (\ref{hyperparameters}), we assumed that male and female participants shared the same expected leniency $\mu_\alpha$ and attractiveness $\mu_\beta$, by setting both of these expectations to zero. We can relax this by allowing the genders to have a different expectation for one of these parameters whilst setting the other to zero. The  distribution for $\mu_\alpha$ and $\mu_\beta$ becomes:
\begin{align}
\left[ \begin{array}{c}
\alpha_a \\
\beta_a \\
\end{array} \right]
& \ \sim \
\textnormal{N}\left(
\left[\begin{array}{c}
m_a\mu_\textnormal{male} \\
0 \\
\end{array} \right]
,
\left[ \begin{array}{cc}
\sigma_\alpha^2 & \rho_{\alpha\beta}\sigma_\alpha\sigma_\beta \\
\rho_{\alpha\beta}\sigma_\alpha\sigma_\beta  & \sigma_\beta^2 
\end{array} \right]
\right). 
\end{align}
Here, $\mu_\textnormal{male}$ is the difference between the expected attractiveness of males and females respectively, and $m_a$ is an indicator for whether individual $a$ is male. 
This gender parameter can also be interpreted as the difference between the expected leniency of females and males.
Hence, a positive $\mu_\textnormal{male}$ would suggest that males were on average more attractive than females, and/or females were more lenient in their ratings of males.
We note that these effects could be disentangled if males rated other males and females rated other females. However, because we do not have such data, $\mu_\textnormal{male}$ can only be interpreted as a linear combination (with unknown constants) of the average additional male attractiveness and average additional female rater leniency.

Estimates are reported under the heading ``with gender'' in the tables.
The gender difference $\mu_\textnormal{male}$ is estimated to be 0.08 with a 95\% credible interval containing zero. 
There is therefore insufficient evidence to suggest a gender difference.
The variance and correlation estimates are virtually the same for the models with and without $\mu_\textnormal{male}$.

\section{Simulations\label{sec:sim}}

We first present the results of a simulation study exploring Bayesian properties of the MCMC estimator for an extended dIRT model that includes a distal outcome.  We generated data for the same data design, size and parameter estimates as in the previous section.  Starting with the estimated values of the variance and correlation hyperparameters, we generated 551 pairs of individual latent traits ($\alpha_a, \beta_a$), and 8,126 directed dyadic latent traits $\gamma_{a, p}$. Using the estimated item step-difficulties from Section \ref{speeddatingemp}, we then generated responses from the dIRT model (\ref{dirt}).  Using the estimated regression coefficients, we finally  generated the distal outcomes according to model (\ref{distal}).
We summarize our findings regarding parameter recovery in the figures below.  

\begin{figure}[H]
\begin{center}
\includegraphics[width=0.45\textwidth,keepaspectratio]{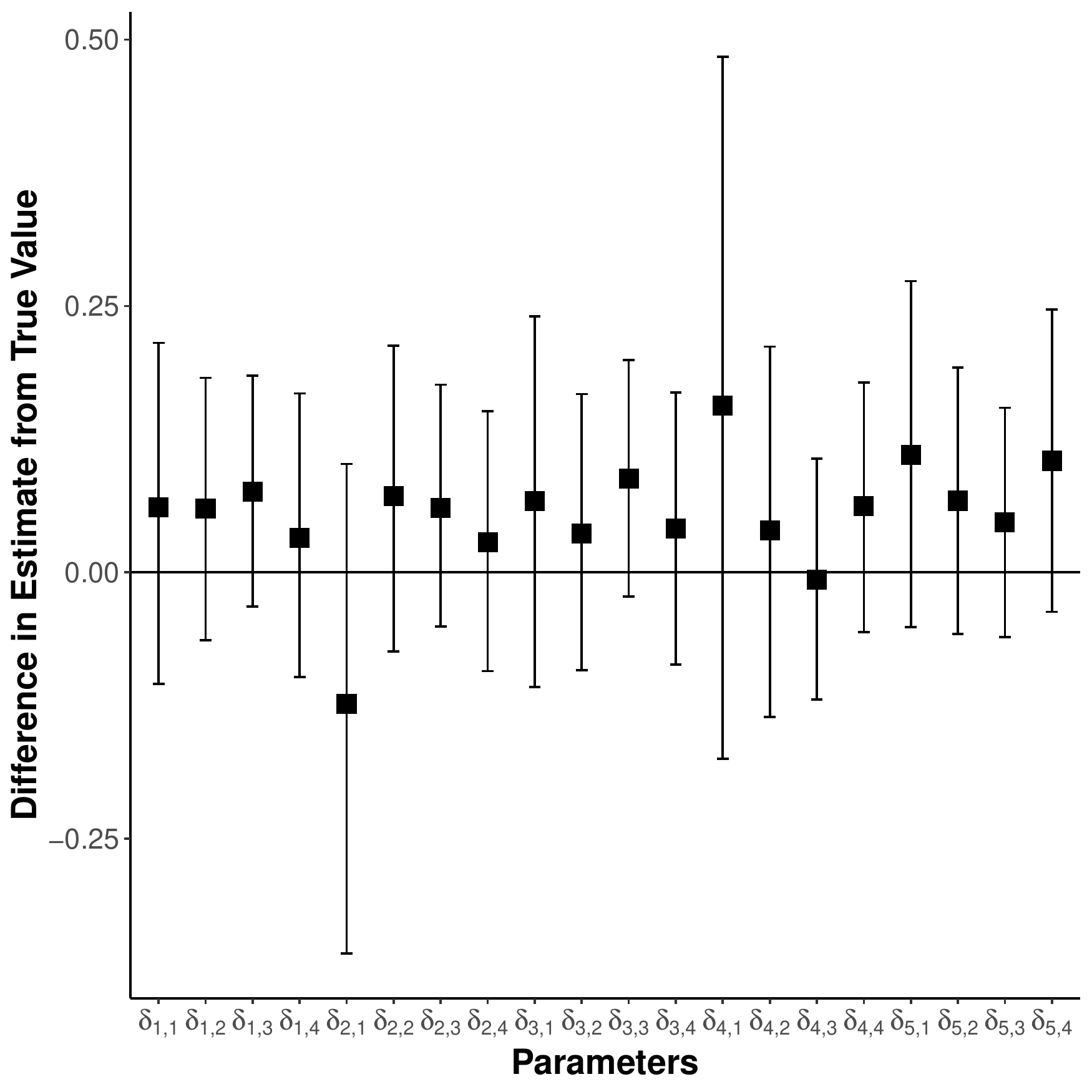}
\caption{Difference between item step difficulty estimates and true parameter values.}
\label{itemrecovery}
\end{center}
\end{figure}

\begin{figure}[h]
\begin{center}
\includegraphics[width=0.45\textwidth,keepaspectratio]{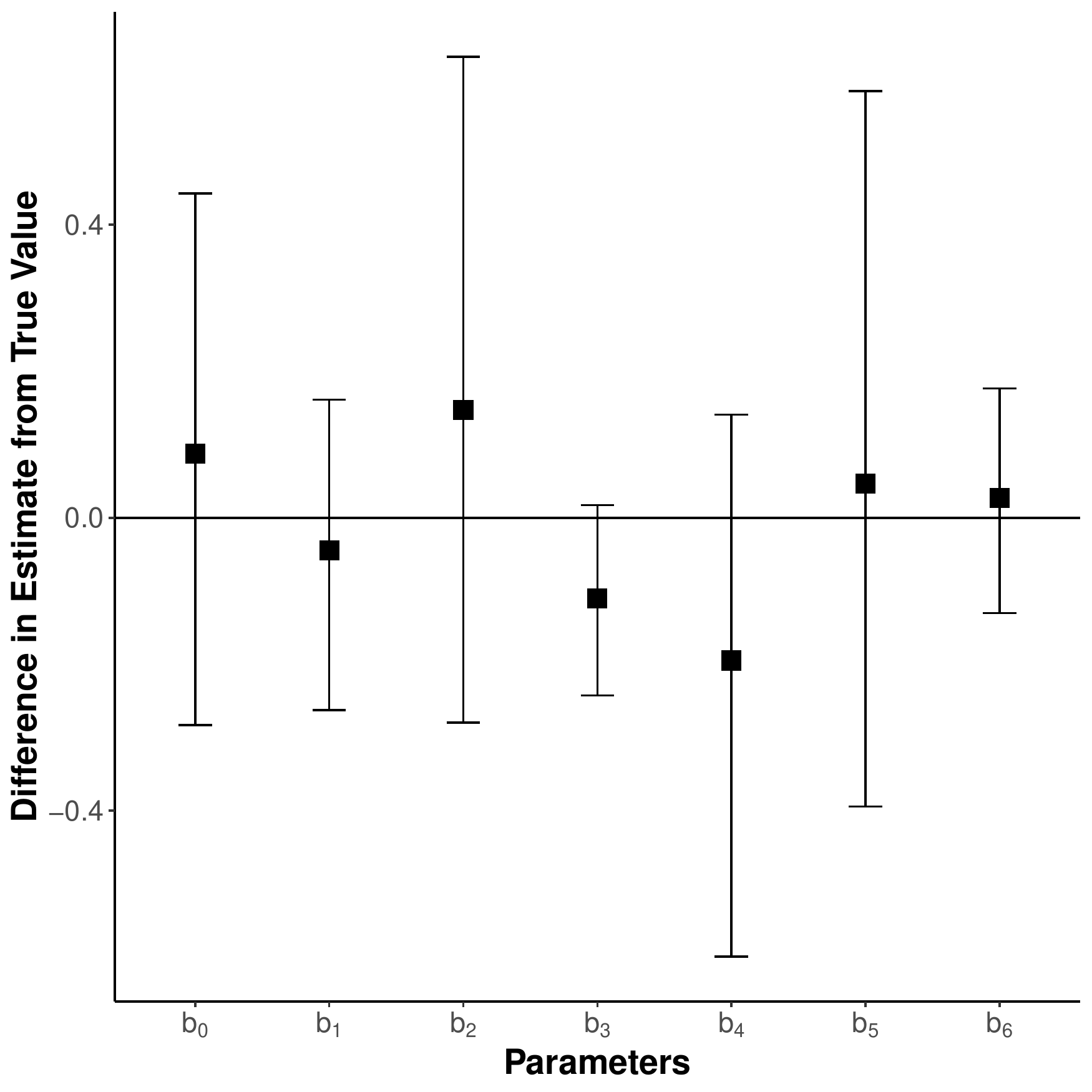}
\caption{Difference between distal outcome regression estimates and  true parameter values.}
\label{distalrecovery}
\end{center}
\end{figure}

Figure \ref{indivrecovery} depicts the difference between the estimated hyperparameters and the actual parameters across all 4,000 draws after convergence.  The square represents the posterior mean of these estimates while the whiskers represent the bounds for the 95\% credible intervals based on the $2.5^{\text{th}}$ and $97.5^{\text{th}}$ quantiles.  Similarly, Figures \ref{itemrecovery} and \ref{distalrecovery} provide the analogous comparison for estimates of the item step parameters and the distal outcome regression parameters, respectively.  We see that all credible intervals contains the true value and our procedure hence has good Bayesian performance. 

\begin{figure}[H]
\begin{center}
\includegraphics[width=0.45\textwidth,keepaspectratio]{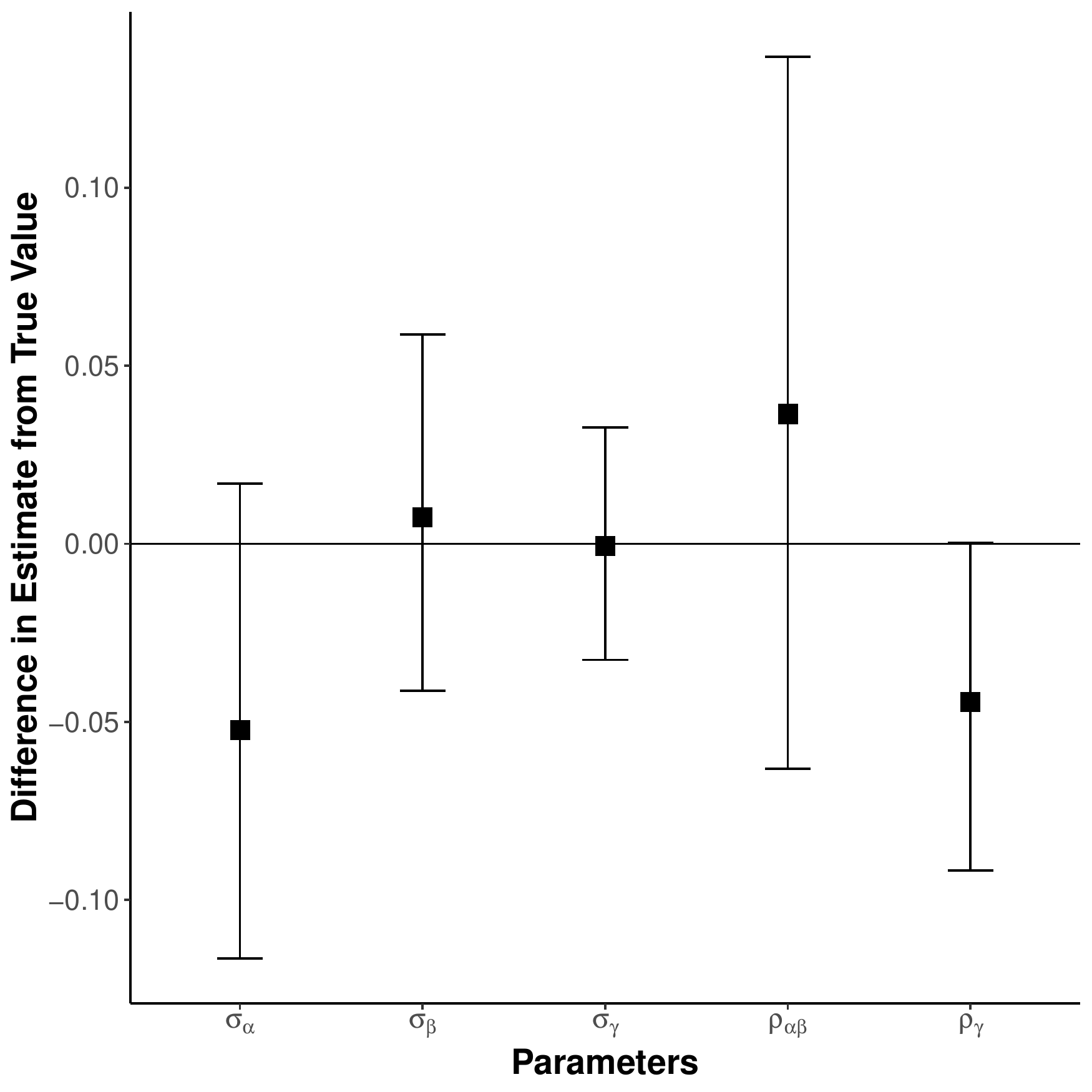}
\caption{Difference between hyperparameter estimates and true parameter values.}
\label{indivrecovery}
\end{center}
\end{figure}

In order to evaluate frequentist properties such as the bias of point estimates and the validity of model-based standard errors, we generated 50 datasets based on the same procedure as above, and estimated the same model for each dataset.  Based on these 50 replications, we then estimated  (i) the  absolute bias of parameter estimates using the difference between the mean (over replications) of the estimated parameters and the true values, and (ii) the relative bias of standard error estimates using the mean (over replications) of the estimated standard errors  divided by the empirical standard deviation (over replications) of the point estimates minus 1. Monte Carlo errors for these quantities were estimated using the formulae in \citet{white:2010}. 

\begin{figure}[h]
\begin{center}
\includegraphics[width=1.0\textwidth,keepaspectratio]{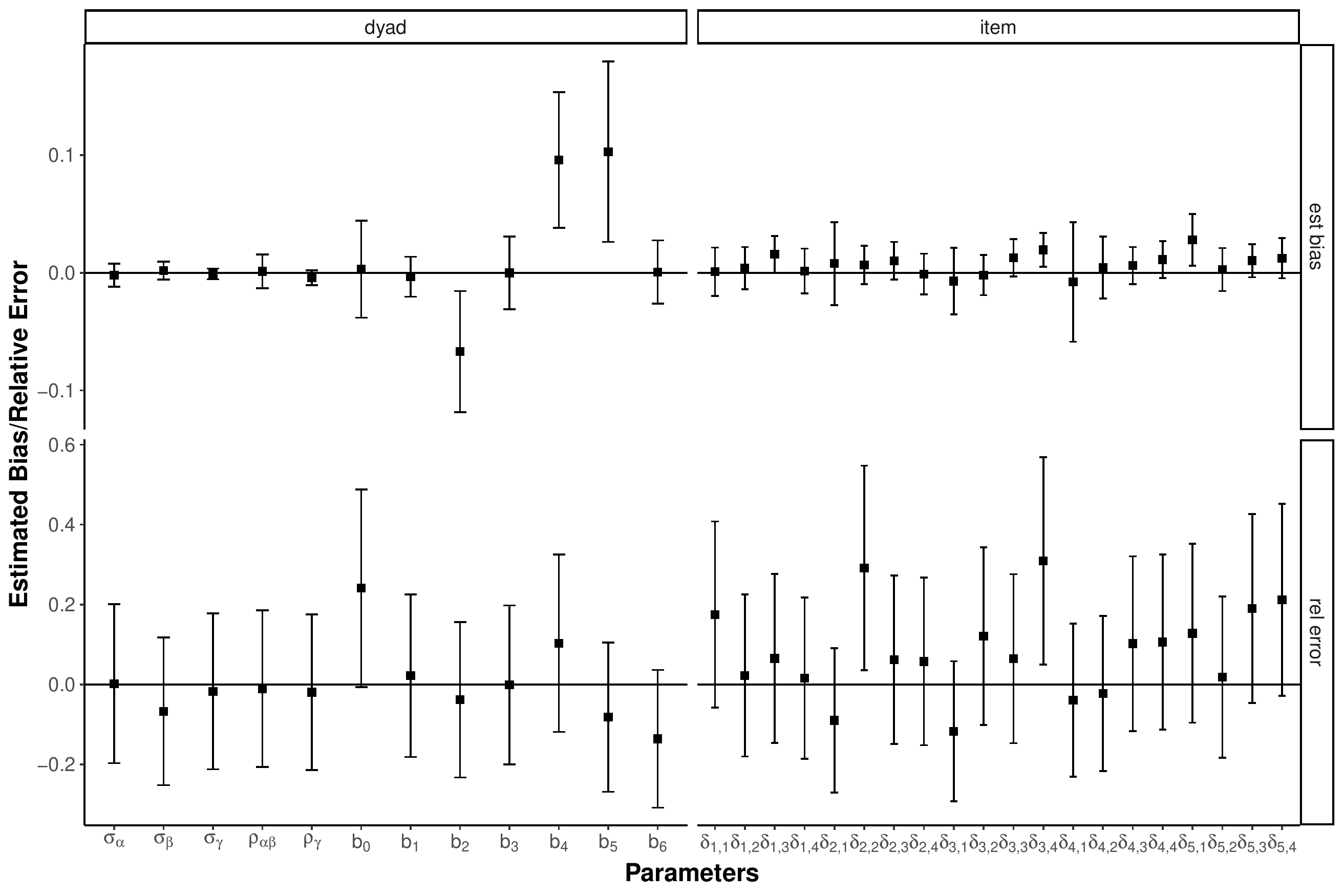}
\caption{Performance of point estimates and standard errors across 50 replications.}
\label{bias}
\end{center}
\end{figure}

In Figure \ref{bias} we show the estimated absolute bias of the parameter estimates (top) and relative error of the standard error estimates (bottom), together with error bars of $\pm 1.96$ times their Monte Carlo error estimates, representing approximate 95\% confidence intervals if the sampling distributions are approximately normal.  
We see that there is small absolute bias in our point estimates across parameters, most of which can be attributed to Monte Carlo error with the exception of $b_2$, $b_4$ and $b_5$.   There is also small relative bias for the standard error estimates, most of which can be attributed to Monte Carlo error with the exception of $\delta_{2,2}$ and $\delta_{3,4}$.
In summary, our procedure has good frequentist properties.

\section{Concluding  Remarks\label{sec:conc}}

We have proposed a dyadic Item Response Theory (dIRT) model that integrates Item Response Theory (IRT) models for measurement and the Social Relations Model (SRM) for dyadic data by modeling the responses of an actor as a function of the actor's inclination to act and the partner's tendency to elicit that action as well as the unique relationship of the pair. We described how the model can be extended to larger group settings,  include covariates for the individual and dyad, include cluster-level random effects, and accommodate distal outcomes.  We also discussed data designs for which the dIRT model is identified, emphasizing that longitudinal data is not required, and described how the model can be estimated using standard software for Bayesian inference. The proposed estimation approach was shown to have good performance in simulation studies. 

The practical utility of the dIRT model was demonstrated by applying it to speed dating data with ordinal items. 
The estimated variance of the actor effect suggests that there was some variation in the way different individuals rated the same sets of partners, or in other words that there was a large variation in how lenient individuals were in rating their partners.   The estimated variance of the partner effect can be thought of as reflecting how attractive the partner is, on average, to all other individuals, and indicates that there is some degree of universal attractiveness. We found that there is evidence for a unique interaction effect (dyadic latent trait) and that the magnitude of this effect helps predict whether the individuals want to see each other again. This finding suggests that the dyadic latent trait has predictive validity, a conclusion that can perhaps be more easily justified when a sequential estimation approach is used. A traditional IRT model, measuring individual latent traits only, would ignore this dyadic latent trait, which can be thought of as the ``eye-of-the-beholder'' effect. The 
dyadic latent traits were positively correlated within dyads, suggesting that both members of a dyad tended to perceive their interaction similarly.

In the speed-dating application, the dyadic latent trait was of particular interest from the point-of-view of matchmaking. In other applications where the actors can be viewed as the raters, ``perceivers''  or informants used to make inferences regarding the partners,  the partner latent trait is of greatest interest. In this case, the advantage of the dIRT is that it purges the measurement of the partner latent trait from both the global rater bias $\alpha$ and the target-specific rater bias $\gamma$. In a collaborative problem-solving task, both the actor and partner latent traits may be of interest, in which case it becomes important to accommodate the dyadic latent trait in the model to prevent it from contaminating the individual latent traits of interest. The dyadic latent trait could in this case be viewed as a nuisance reflecting a fortunate or unfortunate choice of collaborator. For all these types of applications, dyadic designs that permit estimation of the dIRT are essential.

The formulation of the dIRT model, and providing a viable estimation approach for it, provides researchers with the impetus to collect appropriate data for investigating  dyadic interactions or individual latent traits, free from such interaction effects, in a measurement context.  


\section*{Acknowledgements}
Brian Gin and Nicholas Sim are joint first authors of this paper. This work was partly funded by the Research Council of Norway through its Centres of Excellence funding scheme, project number 262700, and by the Institute of Education Sciences, U.S. Department of Education, through Grant R305D140059. 

\bibliographystyle{apa}
\bibliography{mybib} 

\newpage

\section*{Appendix A: \texttt{Stan} Code}
\begin{Verbatim}[fontsize=\fontsize{9}{9}]
# clears workspace: 
rm(list = ls())

library(rstan)
rstan_options(auto_write = TRUE)
options(mc.cores = 8)

library(tidyverse)

# load dataset:
load(file = "df.complete.Rdata")
load(file = "dpair.specific.Rdata")

# no gen with int stan model
modelngwi <- "
functions {
    real pcminteract(int x, real alpha, real beta, real gamma, vector delta) {
      vector[rows(delta) + 1] unsummed;
      vector[rows(delta) + 1] probs;
      unsummed = append_row(rep_vector(0.0, 1), alpha + beta + gamma - delta);
      probs = softmax(cumulative_sum(unsummed));
      return categorical_lpmf(x+1 | probs);
    }
  }
data {
  int<lower = 1> I;                  // # items
  int<lower = 1> A;                  // # actors (or partners)
  int<lower = 1> U;                  // # undirected pairs
  int<lower = 1> N;                  // # responses
  int<lower = 1> D;                  // # decisions
  int<lower = 1> B;                  // integer value for # distal regression parameters
  int<lower = 1, upper = A> aa[N];   // size N array to index actors for each response
  int<lower = 1, upper = A> pp[N];   // size N array to index partners for each response
  int<lower = 1, upper = I> ii[N];   // size N array to index items for each response
  int<lower = 0> x[N];               // size N array for responses; x = 0, 1 ... m_i
  int<lower = 1, upper = U> dd[N];   // size N array to index undirected pairs for each response
  int<lower = 1, upper = 2> mm[N];   // size N array to index match for each response
  int<lower = 1, upper = A> aaa[D];  // size D array to index actors for each decision
  int<lower = 1, upper = A> ppp[D];  // size D array to index partners for each decision
  int<lower = 1, upper = U> ddd[D];  // size D array to index undirected pairs for each decision
  int<lower = 1, upper = 2> mmm[D];  // size D array to index match for each decision
  int<lower = 0, upper = 1> zzz[D];  // size D array for decisions
}
transformed data {
  int M;                             // # parameters per item (same for all items)
    M = max(x);
}
parameters {
  vector[M] delta[I];                // length m vector for each item i
  vector[2] AB[A];                   // size 2 vector of alpha and beta for each person; 
  vector[2] GG[U];                   // size 2 vector of gammas for each undirected pair;
  real<lower = 0> sigmaA;            // real sd of alpha 
  real<lower = 0> sigmaB;            // real sd of beta 
  real<lower = 0> sigmaG;            // real sd of gamma 
  real<lower = -1, upper = 1> rhoAB; // real cor between alpha and beta (within person)
  real<lower = -1, upper = 1> rhoG;  // real cor between gammas (within pair)
  real beta[B];                      // B-dimensional array of real valued of beta 
                                     // (distal regression parameters) 
}
transformed parameters {
  cov_matrix[2] SigmaAB;             // 2x2 covariance matrix of alpha and beta
  cov_matrix[2] SigmaG;              // 2x2 covariance matrix of gammas
  SigmaAB[1, 1] = sigmaA^2;
  SigmaAB[2, 2] = sigmaB^2;
  SigmaAB[1, 2] = rhoAB * sigmaA * sigmaB;
  SigmaAB[2, 1] = rhoAB * sigmaA * sigmaB;
  SigmaG[1, 1] = sigmaG^2;
  SigmaG[2, 2] = sigmaG^2;
  SigmaG[1, 2] = rhoG * sigmaG^2;
  SigmaG[2, 1] = rhoG * sigmaG^2;
}
model {
  AB ~ multi_normal(rep_vector(0.0, 2), SigmaAB);
  GG ~ multi_normal(rep_vector(0.0, 2), SigmaG);
  for (n in 1:N){
    target += pcminteract(x[n], AB[aa[n],1], AB[pp[n],2], GG[dd[n], mm[n]], delta[ii[n]]);
  }
  for (d in 1:D){
    //distal logistic regression
    target += bernoulli_logit_lpmf(zzz[d] | (beta[1] 
    + beta[2]*AB[aaa[d],1] 
    + beta[3]*AB[ppp[d],1] 
    + beta[4]*AB[aaa[d],2]
    + beta[5]*AB[ppp[d],2]  
    + beta[6]*GG[ddd[d], mmm[d]] 
    + beta[7]*GG[ddd[d], (3-mmm[d])] 
    + beta[8]*AB[aaa[d],1]*AB[ppp[d],1] 
    + beta[9]*AB[aaa[d],2]*AB[ppp[d],2]
    + beta[10]*GG[ddd[d], mmm[d]]*GG[ddd[d], (3-mmm[d])]));
  }
}
"

# no gen with int model
I <- max(df.complete$item)
A <- max(df.complete$actor)
U <- max(df.complete$unique.pair)
N <- nrow(df.complete)
D <- nrow(dpair.specific)
B <- 10

data <- list(I = I,
             A = A,
             U = U,
             N = N,
             D = D,
             B = B,
             aa = as.numeric(df.complete$actor),
             pp = as.numeric(df.complete$partner),
             ii = as.numeric(df.complete$item),
             x = as.numeric(df.complete$x),
             dd = as.numeric(df.complete$unique.pair),
             mm = as.numeric(df.complete$selector),
             aaa = as.numeric(dpair.specific$actor),
             ppp = as.numeric(dpair.specific$partner),
             ddd = as.numeric(dpair.specific$unique.pair),
             mmm = as.numeric(dpair.specific$selector),
             zzz = as.numeric(dpair.specific$decision))

set.seed(349)
samples <- stan(model_code=modelngwi,   
                 data=data,
                 iter=2000, 
                 chains=4,
                 seed = 349)

pcm_estimated_values <- summary(samples,
                                pars = c("sigmaA",
                                         "sigmaB",
                                         "sigmaG",
                                         "rhoAB",
                                         "rhoG",
                                         "beta"),
                                probs = c(.025, .975))
View(pcm_estimated_values$summary)

# no gen no int stan model
modelngni <- "
functions {
    real pcminteract(int x, real alpha, real beta, real gamma, vector delta) {
      vector[rows(delta) + 1] unsummed;
      vector[rows(delta) + 1] probs;
      unsummed = append_row(rep_vector(0.0, 1), alpha + beta + gamma - delta);
      probs = softmax(cumulative_sum(unsummed));
      return categorical_lpmf(x+1 | probs);
    }
  }
data {
  int<lower = 1> I;                  // # items
  int<lower = 1> A;                  // # actors (or partners)
  int<lower = 1> U;                  // # undirected pairs
  int<lower = 1> N;                  // # responses
  int<lower = 1> D;                  // # decisions
  int<lower = 1> B;                  // integer value for # distal regression parameters
  int<lower = 1, upper = A> aa[N];   // size N array to index actors for each response
  int<lower = 1, upper = A> pp[N];   // size N array to index partners for each response
  int<lower = 1, upper = I> ii[N];   // size N array to index items for each response
  int<lower = 0> x[N];               // size N array for responses; x = 0, 1 ... m_i
  int<lower = 1, upper = U> dd[N];   // size N array to index undirected pairs for each response
  int<lower = 1, upper = 2> mm[N];   // size N array to index match for each response
  int<lower = 1, upper = A> aaa[D];  // size D array to index actors for each decision
  int<lower = 1, upper = A> ppp[D];  // size D array to index partners for each decision
  int<lower = 1, upper = U> ddd[D];  // size D array to index undirected pairs for each decision
  int<lower = 1, upper = 2> mmm[D];  // size D array to index match for each decision
  int<lower = 0, upper = 1> zzz[D];  // size D array for decisions
}
transformed data {
  int M;                             // # parameters per item (same for all items)
    M = max(x);
}
parameters {
  vector[M] delta[I];                // length m vector for each item i
  vector[2] AB[A];                   // size 2 vector of alpha and beta for each person; 
  vector[2] GG[U];                   // size 2 vector of gammas for each undirected pair;
  real<lower = 0> sigmaA;            // real sd of alpha 
  real<lower = 0> sigmaB;            // real sd of beta 
  real<lower = 0> sigmaG;            // real sd of gamma 
  real<lower = -1, upper = 1> rhoAB; // real cor between alpha and beta (within person)
  real<lower = -1, upper = 1> rhoG;  // real cor between gammas (within pair)
  real beta[B];                      // B-dimensional array of real valued of beta 
                                     // (distal regression parameters) 
}
transformed parameters {
  cov_matrix[2] SigmaAB;             // 2x2 covariance matrix of alpha and beta
  cov_matrix[2] SigmaG;              // 2x2 covariance matrix of gammas
  SigmaAB[1, 1] = sigmaA^2;
  SigmaAB[2, 2] = sigmaB^2;
  SigmaAB[1, 2] = rhoAB * sigmaA * sigmaB;
  SigmaAB[2, 1] = rhoAB * sigmaA * sigmaB;
  SigmaG[1, 1] = sigmaG^2;
  SigmaG[2, 2] = sigmaG^2;
  SigmaG[1, 2] = rhoG * sigmaG^2;
  SigmaG[2, 1] = rhoG * sigmaG^2;
}
model {
  AB ~ multi_normal(rep_vector(0.0, 2), SigmaAB);
  GG ~ multi_normal(rep_vector(0.0, 2), SigmaG);
  for (n in 1:N){
    target += pcminteract(x[n], AB[aa[n],1], AB[pp[n],2], GG[dd[n], mm[n]], delta[ii[n]]);
  }
  for (d in 1:D){
    //distal logistic regression
    target += bernoulli_logit_lpmf(zzz[d] | (beta[1] 
    + beta[2]*AB[aaa[d],1] 
    + beta[3]*AB[ppp[d],1] 
    + beta[4]*AB[aaa[d],2] 
    + beta[5]*AB[ppp[d],2]  
    + beta[6]*GG[ddd[d], mmm[d]] 
    + beta[7]*GG[ddd[d], (3-mmm[d])]));
  }
}
"

# no gen with int model
I <- max(df.complete$item)
A <- max(df.complete$actor)
U <- max(df.complete$unique.pair)
N <- nrow(df.complete)
D <- nrow(dpair.specific)
B <- 7

data <- list(I = I,
             A = A,
             U = U,
             N = N,
             D = D,
             B = B,
             aa = as.numeric(df.complete$actor),
             pp = as.numeric(df.complete$partner),
             ii = as.numeric(df.complete$item),
             x = as.numeric(df.complete$x),
             dd = as.numeric(df.complete$unique.pair),
             mm = as.numeric(df.complete$selector),
             aaa = as.numeric(dpair.specific$actor),
             ppp = as.numeric(dpair.specific$partner),
             ddd = as.numeric(dpair.specific$unique.pair),
             mmm = as.numeric(dpair.specific$selector),
             zzz = as.numeric(dpair.specific$decision))

set.seed(349)
samples <- stan(model_code=modelngni,   
                data=data,
                iter=2000, 
                chains=4,
                seed = 349)

pcm_estimated_values <- summary(samples,
                                pars = c("sigmaA",
                                         "sigmaB",
                                         "sigmaG",
                                         "rhoAB",
                                         "rhoG",
                                         "beta",
                                         "delta"),
                                probs = c(.025, .975))
View(pcm_estimated_values$summary)

# with gen no int stan model
modelwgni <- "
functions {
    real pcminteract(int x, real alpha, real beta, real gamma, vector delta) {
      vector[rows(delta) + 1] unsummed;
      vector[rows(delta) + 1] probs;
      unsummed = append_row(rep_vector(0.0, 1), alpha + beta + gamma - delta);
      probs = softmax(cumulative_sum(unsummed));
      return categorical_lpmf(x+1 | probs);
    }
  }
data {
  int<lower = 1> I;                  // # items
  int<lower = 1> A;                  // # actors (or partners)
  int<lower = 1> U;                  // # undirected pairs
  int<lower = 1> N;                  // # responses
  int<lower = 1> D;                  // # decisions
  int<lower = 1> B;                  // integer value for # distal regression parameters
  int<lower = 1, upper = A> aa[N];   // size N array to index actors for each response
  int<lower = 1, upper = A> pp[N];   // size N array to index partners for each response
  int<lower = 1, upper = I> ii[N];   // size N array to index items for each response
  int<lower = 0> x[N];               // size N array for responses; x = 0, 1 ... m_i
  int<lower = 1, upper = U> dd[N];   // size N array to index undirected pairs for each response
  int<lower = 1, upper = 2> mm[N];   // size N array to index match for each response
  int<lower = 0, upper = 1> gg[N];   // size N array to index gender for each response
  int<lower = 1, upper = A> aaa[D];  // size D array to index actors for each decision
  int<lower = 1, upper = A> ppp[D];  // size D array to index partners for each decision
  int<lower = 1, upper = U> ddd[D];  // size D array to index undirected pairs for each decision
  int<lower = 1, upper = 2> mmm[D];  // size D array to index match for each decision
  int<lower = 0, upper = 1> zzz[D];  // size D array for decisions
}
transformed data {
  int M;                             // # parameters per item (same for all items)
    M = max(x);
}
parameters {
  vector[M] delta[I];                // length m vector for each item i
  vector[2] AB[A];                   // size 2 vector of alpha and beta for each person; 
  vector[2] GG[U];                   // size 2 vector of gammas for each undirected pair;
  real<lower = 0> sigmaA;            // real sd of alpha 
  real<lower = 0> sigmaB;            // real sd of beta 
  real<lower = 0> sigmaG;            // real sd of gamma 
  real<lower = -1, upper = 1> rhoAB; // real cor between alpha and beta (within person)
  real<lower = -1, upper = 1> rhoG;  // real cor between gammas (within pair)
  real mu;                           // real value of mean of theta for males
  real beta[B];                      // B-dimensional array of real valued of beta 
                                     // (distal regression parameters) 
}
transformed parameters {
  cov_matrix[2] SigmaAB;             // 2x2 covariance matrix of alpha and beta
  cov_matrix[2] SigmaG;              // 2x2 covariance matrix of gammas
  SigmaAB[1, 1] = sigmaA^2;
  SigmaAB[2, 2] = sigmaB^2;
  SigmaAB[1, 2] = rhoAB * sigmaA * sigmaB;
  SigmaAB[2, 1] = rhoAB * sigmaA * sigmaB;
  SigmaG[1, 1] = sigmaG^2;
  SigmaG[2, 2] = sigmaG^2;
  SigmaG[1, 2] = rhoG * sigmaG^2;
  SigmaG[2, 1] = rhoG * sigmaG^2;
}
model {
  AB ~ multi_normal(rep_vector(0.0, 2), SigmaAB);
  GG ~ multi_normal(rep_vector(0.0, 2), SigmaG);
  for (n in 1:N){
    target += pcminteract(x[n], AB[aa[n],1] - mu*gg[n], AB[pp[n],2], GG[dd[n], mm[n]], delta[ii[n]]);
  }
  for (d in 1:D){
    //distal logistic regression
    target += bernoulli_logit_lpmf(zzz[d] | (beta[1] 
    + beta[2]*AB[aaa[d],1] 
    + beta[3]*AB[ppp[d],1] 
    + beta[4]*AB[aaa[d],2]
    + beta[5]*AB[ppp[d],2]  
    + beta[6]*GG[ddd[d], mmm[d]] 
    + beta[7]*GG[ddd[d], (3-mmm[d])]));
  }
}
"

# no gen with int model
I <- max(df.complete$item)
A <- max(df.complete$actor)
U <- max(df.complete$unique.pair)
N <- nrow(df.complete)
D <- nrow(dpair.specific)
B <- 7

data <- list(I = I,
             A = A,
             U = U,
             N = N,
             D = D,
             B = B,
             aa = as.numeric(df.complete$actor),
             pp = as.numeric(df.complete$partner),
             ii = as.numeric(df.complete$item),
             x = as.numeric(df.complete$x),
             dd = as.numeric(df.complete$unique.pair),
             mm = as.numeric(df.complete$selector),
             gg = as.numeric(df.complete$male),
             aaa = as.numeric(dpair.specific$actor),
             ppp = as.numeric(dpair.specific$partner),
             ddd = as.numeric(dpair.specific$unique.pair),
             mmm = as.numeric(dpair.specific$selector),
             zzz = as.numeric(dpair.specific$decision))

set.seed(349)
samples <- stan(model_code=modelwgni,   
                data=data,
                iter=2000, 
                chains=4,
                seed = 349)

pcm_estimated_values <- summary(samples,
                                pars = c("sigmaA",
                                         "sigmaB",
                                         "sigmaG",
                                         "rhoAB",
                                         "rhoG",
                                         "mu",
                                         "beta"),
                                probs = c(.025, .975))
View(pcm_estimated_values$summary)
\end{Verbatim}
\newpage

\section*{Appendix B: Sequential Estimation}

\footnotesize
Using the sequential estimation approach with multiple imputation described in Section \ref{joint}, the results of first estimating the dyadic partial credit model (ignoring the distal outcome), and subsequently estimating the distal regression are presented in Tables \ref{varcorestseq} and \ref{distalestseq}.  
We report estimates as means of draws, and the values in parentheses represent the $2.5^{\text{th}}$ and the $97.5^{\text{th}}$ quantiles of the parameter estimates.

MCMC estimates for the standard deviations and correlations of the individual and dyadic latent traits are shown in Table \ref{varcorestseq}.  The estimates are qualitatively similar to the estimates from  the joint approach reported in Table \ref{varcorest}.

Estimates for the distal regression based on multiple draws of the latent traits from their posterior distribution are shown in Table \ref{distalestseq}. While the sign of the coefficient estimates are the same as for the joint approach in Table \ref{distalest}, their magnitudes differ substantially.  Overall, the estimates using the sequential approach are smaller in absolute value, particularly for $b_3, b_4$ and $b_5$. This may be because the joint approach effectively treats the distal outcome as an item in the measurement model, and therefore makes the latent traits highly predictive of the distal outcome.

\begin{table}[H]
\caption{Sequential Estimation Approach: Estimates of Standard Deviations and Correlations of Individual and Dyadic Latent Traits}
\label{varcorestseq}
\begin{center}
\scalebox{0.75}{
\begin{tabular}{|l|.,|.,|.,|}
\hline
\multirow{ 2}{*}{}   & \multicolumn{4}{|c|}{without gender}      & \multicolumn{2}{|c|}{with gender}\\ \cline{2-7}
                     & \multicolumn{2}{|c|}{with interactions}   & \multicolumn{2}{|c|}{without interactions}         & \multicolumn{2}{|c|}{without interactions}\\
\hline
$\mu_{\text{male}}$  &         &                 &         &                & $-0.15$ & $(-0.36,0.06)$\\
$\sigma_\alpha$      & $ 1.05$ & $( 0.98,1.13)$  & $ 1.05$ & $( 0.98,1.13)$ & $ 1.05$ & $( 0.98,1.13)$\\
$\sigma_\beta$       & $ 0.71$ & $( 0.66,0.76)$  & $ 0.71$ & $( 0.66,0.76)$ & $ 0.71$ & $( 0.66,0.76)$\\
$\sigma_\gamma$      & $ 0.89$ & $( 0.86,0.92)$  & $ 0.89$ & $( 0.86,0.92)$ & $ 0.89$ & $( 0.86,0.91)$\\
$\rho_{\alpha\beta}$ & $ 0.03$ & $(-0.06,0.13)$  & $ 0.03$ & $(-0.06,0.13)$ & $ 0.04$ & $(-0.06,0.13)$\\
$\rho_\gamma$        & $ 0.35$ & $( 0.30,0.40)$  & $ 0.35$ & $( 0.30,0.40)$ & $ 0.35$ & $( 0.30,0.40)$\\
\hline
\end{tabular}
}
\end{center}
\end{table}

\begin{table}[H]
\caption{Sequential Estimation Approach: Estimates for Distal Outcome Regression}
\label{distalestseq}
\begin{center}
\scalebox{0.8}{
\begin{tabular}{|l|.,|.,|.,|}
\hline
\multirow{ 2}{*}{}   & \multicolumn{4}{|c|}{without gender}      & \multicolumn{2}{|c|}{with gender}\\ \cline{2-7}
                     & \multicolumn{2}{|c|}{with interactions}   & \multicolumn{2}{|c|}{without interactions}         & \multicolumn{2}{|c|}{without interactions}\\
\hline
$b_0$ & $-0.36$ & $(-0.43,-0.29)$ & $-0.36$ & $(-0.43,-0.29)$ & $-0.36$ & $(-0.43,-0.28)$ \\
$b_1$ & $ 0.40$ & $( 0.36, 0.44)$ & $ 0.40$ & $( 0.36, 0.44)$ & $ 0.38$ & $( 0.34, 0.43)$ \\
$b_2$ & $-0.03$ & $(-0.07, 0.00)$ & $-0.03$ & $(-0.07, 0.00)$ & $-0.02$ & $(-0.06, 0.02)$ \\
$b_3$ & $-0.35$ & $(-0.44,-0.26)$ & $-0.34$ & $(-0.42,-0.25)$ & $-0.32$ & $(-0.42,-0.23)$ \\
$b_4$ & $ 1.29$ & $( 1.19, 1.38)$ & $ 1.28$ & $( 1.19, 1.37)$ & $ 1.26$ & $( 1.16, 1.36)$ \\
$b_5$ & $ 0.82$ & $( 0.76, 0.89)$ & $ 0.82$ & $( 0.77, 0.89)$ & $ 0.82$ & $( 0.76, 0.88)$ \\
$b_6$ & $ 0.06$ & $( 0.01, 0.11)$ & $ 0.06$ & $( 0.01, 0.11)$ & $ 0.06$ & $(-0.01, 0.11)$ \\
$b_7$ & $-0.01$ & $(-0.04, 0.01)$ & $     $ & $             $ & $     $ & $             $ \\
$b_8$ & $ 0.18$ & $( 0.09, 0.28)$ & $     $ & $             $ & $     $ & $             $ \\
$b_9$ & $-0.02$ & $(-0.07, 0.04)$ & $     $ & $             $ & $     $ & $             $ \\
\hline
\end{tabular}
}
\end{center}
\end{table}

\end{document}